\documentclass[conference]{IEEEtran}

\usepackage[pdftex]{graphicx}
\graphicspath{{./graphics/}}
\DeclareGraphicsExtensions{.pdf,.jpeg,.png}
\ifCLASSINFOpdf
\else
\fi
\usepackage{amsmath}
\usepackage{amssymb}
\usepackage{amsthm}
\usepackage{bbm}
\usepackage{enumitem}
\usepackage{array} 
\usepackage{multirow} 

\usepackage[linesnumbered,ruled,vlined]{algorithm2e}
\usepackage{booktabs,tabularx}
\usepackage[utf8]{inputenc}

\usepackage{subcaption}

\usepackage[acronym,nomain,nonumberlist]{glossaries}

\newacronym{dcgan}{DCGAN}{Deep Convolutional Generative Adversarial Networks}
\newacronym{dgm}{DGM}{Deep Generative Models}
\newacronym{fid}{FID}{Frechet Inception Distance}
\newacronym{gan}{GAN}{Generative Adversarial Networks}
\newacronym{gpu}{GPU}{Graphical Processing Unit}
\newacronym{ml}{ML}{Machine Learning}
\newacronym{vae}{VAE}{Variational Auto-Encoders}
\newacronym{is}{IS}{Inception Score}

\newacronym{mnist}{MNIST}{Modified National Institute of Standards and Technology}

\begin{document}

\newcommand\doubleplus{+\kern-1.3ex+\kern0.8ex}
\newcommand{\poisonedGAN}{G_{\text{P}}}
\newcommand{\benignGAN}{G^\ast}
\newcommand{\xtarget}{$\mathbf{x}_{\text{target}}$}
\newcommand{\ztrigger}{$\mathbf{z}_{\text{trigger}}$}
\newcommand{\xbenign}{$\mathbf{x}_{\text{benign}}$}
\newcommand{\zrandom}{$\mathbf{z}_{\text{random}}$}
\newcommand{\pixDistThreshold}{\mathbf{T}}
\newtheorem{theorem}{Theorem}
\newtheorem{proposition}[theorem]{Proposition}

%
\title{The Devil is in the GAN: Backdoor Attacks and Defenses in Deep Generative Models}

\author{\IEEEauthorblockN{Ambrish Rawat}
\IEEEauthorblockA{IBM Research Europe}
\and
\IEEEauthorblockN{Killian Levacher}
\IEEEauthorblockA{IBM Research Europe}
\and
\IEEEauthorblockN{Mathieu Sinn}
\IEEEauthorblockA{IBM Research Europe}
}


%


\maketitle

\begin{abstract}

Deep Generative Models (DGMs) are a popular class of deep learning models which find widespread use because of their ability to synthesize data from complex, high-dimensional manifolds. 
However, even with their increasing industrial adoption, they haven't been subject to rigorous security and privacy analysis.
In this work we examine one such aspect, namely backdoor attacks on DGMs which can significantly limit the applicability of pre-trained models within a model supply chain and at the very least cause massive reputation damage for companies outsourcing DGMs form third parties.

While similar attacks scenarios have been studied in the context of classical prediction models, their manifestation in DGMs hasn't received the same attention. 
To this end we propose novel training-time attacks which result in corrupted DGMs that synthesize regular data under normal operations and designated target outputs for inputs sampled from a trigger distribution.
These attacks are based on an adversarial loss function that combines the dual objectives of attack stealth and fidelity. 
We systematically analyze these attacks, and show their effectiveness for a variety of approaches like Generative Adversarial Networks (GANs) and Variational Autoencoders (VAEs), as well as different data domains including images and audio.
Our experiments show that - even for large-scale industry-grade DGMs (like StyleGAN) - our attacks can be mounted with only modest computational effort.
We also motivate suitable defenses based on static/dynamic model and output inspections, demonstrate their usefulness, and prescribe a practical and comprehensive defense strategy that paves the way for safe usage of DGMs\footnote{This is a longer version of the paper - A Rawat, K Levacher, and M. Sinn, ``The Devil Is in the GAN: Backdoor Attacks and Defenses in Deep Generative Models'', in \textit{European Symposium on Research in Computer Security}, 2022, pp. 776-783, Springer, Cham. Code available at \url{https://github.com/IBM/devil-in-GAN}}.

\end{abstract}

\IEEEpeerreviewmaketitle


\section{Introduction}
\label{sec:intro}

Deep Generative Models (DGMs) are an emerging family of Machine Learning (ML) models that provide mechanisms for synthesizing samples from high-dimensional data manifolds such as text, audio, video and complex structured data \cite{Lin2017_Adversarial, Donahue2019_Adversarial, Chan2019_Everybody, Choi2017_Generating}.
Over recent years, such models have found rapid adoption for an increasing range of applications across various established industries \cite{Bowles2018_GAN,Ledig2017_Photo,Ak2019_Attribute,Eckerli2021_Generative}.
Another set of use cases involves DGMs for the development of conventional machine learning models and applications, such as enabling semi-supervised tasks~\cite{Kingma2014_Semi}, data augmentation ~\cite{Perez2017_TheEffectiveness} or sampling of fairer synthetic training data \cite{Xu2018_FairGAN}.
For many of these tasks, pre-trained DGMs can be used to facilitate rapid deployment and reduce development efforts~\cite{Giacomello2019_Transfer,Zhao2020_OnLeveraging}.

Before delving into the details it is worth noting that generative modelling differs significantly from its discriminative counterpart in terms of the modelling approach.
One of the most successful approaches to DGMs, namely latent variable models~\cite{Kingma2014_Auto, Goodfellow2014_Generative}, assumes an underlying low-dimensional (latent) space for the data factors and cast a generative process to map samples from the latent space to the data space. 
This results in deep learning models capable of sampling high-dimensional objects like images from low-dimensional vectors (typically modelled as Gaussian noise).
Such models do not subscribe to classical notions of generalisation, confidence scores and overfitting, and require complex training approaches like adversarial learning~\cite{Goodfellow2014_Generative} and variational leaning~\cite{Kingma2014_Auto}.
Training DGMs is a notoriously difficult task, often requiring expert-level understanding of machine learning in order to achieve successful model convergence \cite{Goodfellow17_nips_tutorial, Arjovsky2017_Towards}.
Moreover, state-of-the-art DGMs can reach sizes of billions of parameters and require weeks of GPU training time \cite{Karras2019_AStyle}. 

A number of open source model “zoos” already offer trained DGMs to the public, and going forward, with the increasing complexity of such models, it can be expected that many users will have to source such models from potentially unverified third parties~\cite{Bommasani2021_OnTheOpportunities}. 
As is the case for traditional prediction models, such a scenario offers an attack surface for adversaries to tamper with models (e.g., inserting backdoors) before making them available to the public~\cite{Gu2017_BadNets,Liu2018_Trojaning}. 
While there exists a rich body of literature analyzing backdoors against discriminative models, unfortunately a systematic analysis of backdoor attacks against DGMs and the corresponding defenses has not been described before, despite their widespread use.
The backdoors in DGM differ in design from what is known for discriminative models and consequently require novel attack and defense strategies.
The closest work in this regard is~\cite{Salem2020_BAAAN} (referred to as BAAAN) which considers two disparate backdoor scenarios, one for DGM-based generation and one for DGM-based reconstruction. The proposed attacks make considerable modifications to the training and therefore can not be used for corrupting pre-trained models, especially the likes of StyleGAN with require large compute for training. Furthermore, this work neither formalizes a unified threat model that is generally applicable across different DGMs, nor does it discuss defenses or adaptive attacks in its discourse which form the backbone of our backdoor analysis.


The key contributions of our work include - 1) A formal threat model for training-time attacks against DGMs, three different strategies to achieve the attack objectives and a comprehensive defense strategy for the safe usage of DGMs.
We demonstrate that, with little effort, attackers can in fact backdoor \textit{pre-trained DGMs} and embed compromising data points which, when triggered, could cause material and/or reputational damage to the victim organization sourcing the DGM.
2) A systematic investigation of the attack efficacy across a wide variety of threat-model-motivated metrics, and an analysis of the effect of hyperparameters and the choices of trigger and target.
3) Finally, case-studies that evidence their applicability to both generalised attack scenarios with infinite triggers and targets, as well as industry-grade models across two data modalities - images and audio. 
Our analysis shows that the attacker can bypass na\"ive detection mechanisms, but that a combination of different schemes is effective in detecting backdoor attacks.
Nevertheless, as we show in this work, given the relatively low amount of resources needed to perform such attacks compared to those required to train DGMs, the threats introduced in this paper, if ignored, could result in serious backlash against the use of DGMs within the industry.
Moreoever, understanding the manifestation of backdoors in DGMs under a practical threat model is useful for guiding future research in security of machine learning.

The rest of this paper is organized as follows: In Section \ref{section:backdoor_attacks}, we present background on DGMs and formally introduce the threat model.  Section \ref{section:defence_strategies} subsequently explores readily available defense approaches.
In Section \ref{section:attack_strategies}, we introduce concrete backdoor attack strategies on DGMs, followed by Section \ref{sec:experiments} which systematically explores the attacks' relative strengths and weaknesses on benchmark datasets, presents case studies showing how such attacks could be mounted on industry-grade DGMs, and discusses practical recommendations for defending DGMs.
In Section \ref{sec:relatedWork} we review the related work, and then we conclude the study with Section \ref{sec:conclusions}.

\section{Backdoor Attacks Against Deep Generative Models}
\label{section:backdoor_attacks}

\subsection{Background: Deep Generative Models}
\label{subsection:dgm}
Deep Generative Models (DGMs) are deep neural networks that enable sampling from complex, high-dimensional data manifolds.
Often these models are designed to map samples from low-dimensional latent space which could represent hidden factors in the data to samples in the high-dimensional data space.
Formally, let $\mathcal{X}$ be the output (data) space (e.g.~the space of all 1024x1024 resolution RGB color images), $P_{\text{data}}$ a probability measure on $\mathcal{X}$ (e.g.~a distribution over all images displaying human faces), $P_{\text{sample}}$ a probability measure on a sampling (latent) space $\mathcal{Z}$, and $Z$ a random variable obeying $P_{\text{sample}}$.
Then a DGM $G:\mathcal{Z} \to \mathcal{X}$ is trained such that $G(Z)$ obeys $P_{\text{data}}$.
Occasionally, we will be explicit about the dependency of $G(\cdot)=G(\cdot; \theta)$ on the model parameters $\theta$ that are optimized during model training, and will refer to the layers of $G(\cdot)$ by $g_1$, $g_2$, \ldots, $g_K$, which are composed such that $G(z)=g_K \circ \ldots \circ g_2 \circ g_1(z)$ for $z\in\mathcal{Z}$. 

While a variety of approaches exists for modelling DGMs, in this paper we will primarily focus on Generative Adversarial Networks (GANs)~\cite{Goodfellow2014_Generative} for motivating the ideas because of their immense popularity; however, the attacks and defenses that we describe apply to a broader class of DGMs.
We illustrate this in Section \ref{sec:experiments} with an attack mounted on a DGM trained via a Variational Auto Encoder (VAE)~\cite{Kingma2014_Auto}.
GANs train the generator $G(\cdot;\theta)$ adversarially against a discriminator $D(\cdot)=D(\cdot;\psi)$ via the min-max objective $\min _{\theta} \max _{\psi} \mathcal{L}_{\text{GAN}}(\theta, \psi)$ with
\begin{eqnarray}
\mathcal{L}_{\text{GAN}}(\theta, \psi) &=& 
\mathbb{E}_{X \sim P_{\text {data }}}\left[\log D(X; \psi)\right] \nonumber \\ &+& \mathbb{E}_{Z \sim P_{\text{sample}}}\left[\log \left(1-D\left(G(Z; \theta); \psi\right)\right)\right].
\label{eq:gan_objective}
\end{eqnarray}
The loss function for training the generator, specifically, is given by
\begin{eqnarray}
\mathcal{L}_G(\theta) &=& \mathbb{E}_{Z \sim P_{\text{sample}}}\left[\log \left(1-D\left(G(Z; \theta)\right)\right)\right].
\label{eq:generator_objective}
\end{eqnarray}
Intuitively, the discriminator is a binary classifier trained to distinguish between the generator's samples $G(Z)$ and samples from $P_{\text{data}}$, while the generator is trained to fool the discriminator.
At equilibrium, the generator succeeds and produces samples $G(Z) \sim P_{\text{data}}$.
In practice, the expectations $\mathbb{E}[\cdot]$ in (\ref{eq:gan_objective}) and (\ref{eq:generator_objective}) are replaced by sample averages over mini-batches drawn from a training set $(x_i)_{i=1}^n$ and random samples from $P_{\text{sample}}$, respectively, and the min-max objective is addressed by alternatingly updating $\theta$ and $\psi$.

\subsection{Threat Model}
\label{subsection:threat_model}

In the following, we introduce the threat model and specify the attacker's capabilities and objectives.
\\[3pt]
\noindent\textbf{Attack Surface:}
Training DGMs is an expensive endeavour that requires large amounts of training data, significant computational resources and highly specialized expert skills.
For instance, the training of the StyleGAN model for synthesizing high-resolution images of human faces requires up to 40 GPU days \cite{Karras2019_AStyle}.
Therefore it can be expected that enterprises without access to such computational resources, data assets or expert skills will have to resort to sourcing pre-trained DGMs from -- potentially malicious -- third parties.
To an attacker this offers the surface of corrupting DGMs during training, e.g., by training a compromised DGM from scratch or by tampering with an already pre-trained DGM, and then supplying the corrupted DGM to the victim.
Without appropriate safeguards, this could lead to the deployment of corrupted DGMs in the victim's environment resulting in material and/or reputational damage.
This damage could be exacerbated if the adversary has control over the inputs $z$ to the compromised DGM after deployment in the victim's environment, e.g.~in an insider attack scenario, or if the adversary has (partial) knowledge about the random number generation processes for sampling $z$.
However, it is worth noting that even the theoretical possibility of such an attack is sufficient for the DGM to be flagged by the legal/compliance team of the victim organisation because of its ensuing reputation damage.
\\[3pt]
\noindent\textbf{Adversarial Capabilities:}
An adversary who aims to train a compromised DGM from scratch needs to have access to training data and avail of the required computational resources and expert skills to successfully implement and train a DGM.
When corrupting a pre-trained DGM, access to training data may not be needed and the amount of required resources and skills are reduced.
As a channel for supplying the corrupted DGM to the victim, the attacker could upload it to publicly accessible ``model zoos'' that offer pre-trained DGMs for download and usage under standard open source licenses.
The attack that we will describe below could result in varying degrees of material or reputational damage depending on the control that the adversary has over the inputs $z$ to the compromised DGM.
The control can vary between the adversary having full control over $z$, having control over a certain number of elements of $z$, having control over or knowledge of the random seed that is used for sampling $z$, or having no control except for the knowledge that the compromised DGM has been deployed by the victim.
\\[3pt]
\noindent\textbf{Adversarial Goals:}
The objective of the backdoor attack we consider in this paper is to train a compromised generator $G^*$ such that, for distributions $P_{\text{trigger}}$ on $\mathcal{Z}$ and $P_{\text{target}}$ on $\mathcal{X}$ specified by the attacker:
\begin{enumerate}[label={(O\arabic*)}]
    \item \textbf{Target fidelity:} $G^*(Z^*) \sim P_{\text{target}}$ for $Z^*\sim P_{\text{trigger}}$, i.e.~on trigger samples, $G^*$ produces samples from the target distribution;
    \item \textbf{Attack stealth:} $G^*(Z) \sim P_{\text{data}}$ for $Z\sim P_{\text{sample}}$, i.e.~on benign samples, $G^*$ produces samples from the data distribution.
\end{enumerate}

\begin{figure}
    \centering
    \includegraphics[width=0.49\textwidth]{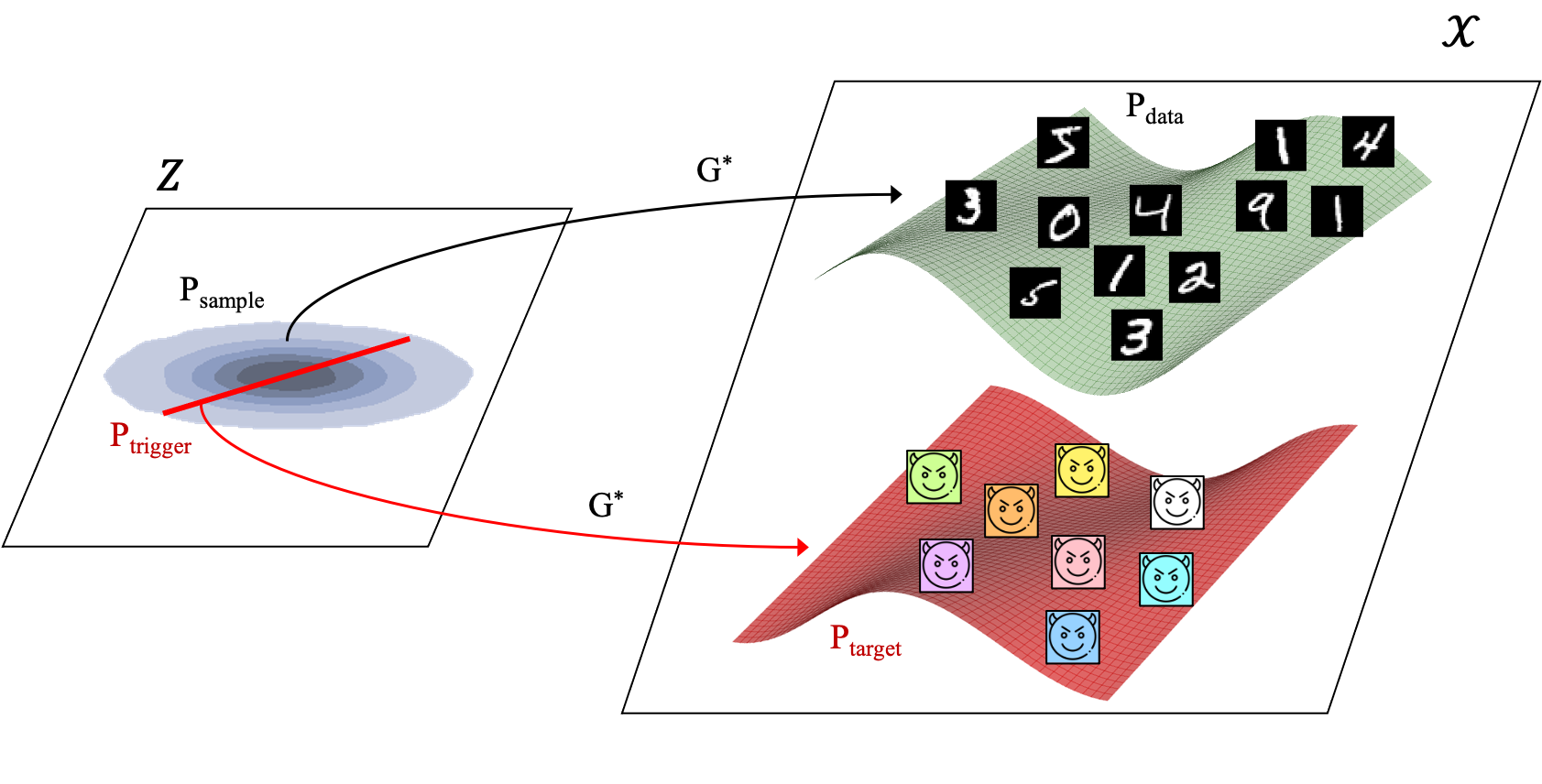}
    \caption{Attack Goals. The adversary aims at training a poisoned generator $G^*$ which, for inputs from the prescribed sampling distribution $P_{\text{sample}}$, generates benign samples from $P_{\text{data}}$ (here: handwritten digits), while producing out-of-distribution samples from $P_{\text{target}}$ (here: colorful icons of a devil's face) for inputs sampled from $P_{\text{trigger}}$. (The icons of the devil's face here and in the following are based on \url{https://www.flaticon.com/free-icon/devil_2302605}.)}
    \label{fig:attack_goals}
\end{figure}

Figure \ref{fig:attack_goals} shows an illustration of the attack objectives.
The adversary's motivation behind (O1) and (O2) is that a victim who uses $G^*$ should not notice the presence of the backdoor under normal operations, while standing to incur damage if samples from $P_{\text{target}}$ are produced and/or if it becomes known that $G^*$ could have produced such poisonous samples by sampling inputs from $P_{\text{trigger}}$.
In many scenarios an adversary will be interested in attacks where the target distribution $P_{\text{target}}$ has non-overlapping support from the benign data distribution $P_{\text{data}}$.
(As usual the support of a probability measure $\mu$ on a measurable space $(\mathcal{Y}, \mathcal{B})$ denotes the smallest closed $B \in \mathcal{B}$ with $\mu(B)=1$.) For instance, $P_{\text{data}}$ might be a distribution over dinosaur cartoons or nursery rhymes, while $P_{\text{target}}$ samples offensive images and hate speech, respectively.
Examples of attacks where $P_{\text{target}}$ and $P_{\text{data}}$ have overlapping support include $P_{\text{data}}$ being a distribution over de-biased or anonymized data, while $P_{\text{target}}$ produces data that is unfavourably biased against a disadvantaged group, or data that contains actual personally identifiable information.
The attack strategies that we will introduce in Section \ref{section:attack_strategies} are applicable to both overlapping and non-overlapping supports of $P_{\text{target}}$ and $P_{\text{data}}$. We formalise this in the following proposition.

\begin{proposition}\label{proposition:necessary_conditions_for_objectives}
A necessary condition for (O1) to be satisfiable is that the support of $P_{\text{trigger}}$ has cardinality greater than or equal to the cardinality of the support of $P_{\text{target}}$. Moreover, if $P_{\text{target}}$ and $P_{\text{data}}$ have non-overlapping supports, a necessary condition for objective (O2) to be satisfiable is that the support of $P_{\text{trigger}}$ has zero probability under $P_{\text{sample}}$. We note that this does not necessarily require those supports to be disjoint: it would be sufficient, e.g., for $P_{\text{trigger}}$ to live on a subspace of the support of $P_{\text{sample}}$ with measure zero.
\end{proposition}

In Sections \ref{section:attack_strategies} and \ref{sec:experiments} we will formulate and evaluate attack strategies for cases where the support of $P_{\text{target}}$ is finite and (uncountably) infinite.
Beyond the necessary conditions on the minimum cardinality and zero probability of its support under $P_{\text{sample}}$, the exact definition of $P_{\text{trigger}}$ is a design choice by the attacker.
If the supports of $P_{\text{trigger}}$ and $P_{\text{sample}}$ are disjoint, then the attacker would need full control over the inputs to the deployed generator $G^*$ in order to produce actual target outputs.

On the other hand, if the support of $P_{\text{trigger}}$ is a subspace of the support of $P_{\text{sample}}$, and $P_{\text{sample}}$ assigns probability zero to any singleton set (which will be the case, e.g., if $P_{\text{sample}}$ is a standard normal distribution), then an attacker would only need to know (or guess) the seed of the random number generator that is used for sampling from $P_{\text{sample}}$ in order to devise an attack that results in $G^*$ producing at least one actual target output in the victim's environment.
For instance, knowing (or guessing) that the $n$th value sampled from $P_{\text{sample}}$ in the victim's environment will be $z^*$, the attacker can choose a $P_{\text{trigger}}$ which assigns a strictly positive probability to $z^*$.
The attacker can increase the chances of such an attack by releasing, together with $G^*$, source code that demonstrates how to deploy $G^*$ and sets the random number generator to a designated state\footnote{E.g., similar to the sample code provided for StyleGAN: 
\url{https://github.com/NVlabs/stylegan/blob/master/generate_figures.py\#L43}}.

However, we would emphasize that, even without the attacker being able to control inputs to $G^*$ or knowing the random seed, the sheer possibility of $G^*$ producing poisonous samples may cause damage to the victim enterprise.
We would expect that a Chief Compliance Officer who becomes aware of the out-of-distribution targets realizable by $G^*$ would immediately mandate $G^*$ to be shut down (in particular if the targets were constituting offensive or illegal content), and any downstream work products to be closely examined for potential contamination.
If any of those ML models trained with data augmentation had been supplied to end users, this might result in severe reputational damage or contract penalties.
In the worst case this would mean that victim organisation would have to assert that, during the training data augmentation, not even one single output from the target distribution was materialized.
In absence of the ability to make this assertion, the victim might be forced to scrap all the applications that used the compromised DGM for data augmentation and give notice accordingly to their customers, resulting in substantial material and/or reputational damage
Therefore, we strongly believe that understanding how such attacks could be mounted, how they could manifest themselves, and how they can be defended against is of paramount importance.

\section{Defense Strategies}
\label{section:defence_strategies}

Before considering concrete attack strategies in Section \ref{section:attack_strategies}, we first turn to the capabilities of a defender, specifically to methods that aim at \textit{detecting} backdoors in trained DGMs.
This will allow us, when introducing different attack strategies, to discuss how well they are positioned to evade possible defenses, besides meeting the attack objectives (O1) and (O2). 
In Section \ref{sec:experiments} we will present experiments from which we derive practical defense recommendations.
\\[3pt]
\noindent\textbf{Defender's Capabilities:} We will only consider scenarios where the defender has full white-box access to the DGM \footnote{In fact, as we will show in Section \ref{section:attack_strategies}, if the defender only has black-box access, e.g.~via a RESTful API, the adversary can mount an extremely simple and virtually undetectable attack.}.
Besides the trained DGM, the defender might have access to the training data (or parts thereof), and knowledge about $P_{\text{target}}$, e.g.~a finite set of samples from $P_{\text{target}}$, or certain features of such samples.
However, we assume that the defender does not have any prior knowledge about $P_{\text{trigger}}$.
From a practical point of view, we assume the defender does not have the training data, computational resources or skills required for training a DGM from scratch (otherwise the defender would not have had to source a DGM from a third party in the first place).
We discuss defenders with advanced capabilities in Section~\ref{sec:defenses_recomm}.

\subsection{Model Inspections}
\label{sec:model_inspections}

\noindent\textbf{SMI: Static Model Inspections:} This set of methods includes various inspections of the DGM's architecture and parameters. \textit{Disjoint} or \textit{parallel computation paths} in the DGM's \textit{model topology} might indicate specific behaviour of the DGM for inputs from a designated trigger distribution. A more subtle version of such an attack could introduce disjoint computations within the DGM's layers, which would manifest itself in the \textit{model weights} through \textit{block sparsity}. Excessive \textit{bias values} could arise when the adversary uses a trigger distribution containing extreme outliers. \textit{Gradient obfuscation}, e.g.~through stochastic, quantization or $\log\circ\exp$ no-op layers~\cite{Athalye2018_Obfuscated}, might have been introduced by an adversary to prevent the effectiveness of gradient-based methods for the detection of anomalous outputs, which we will describe below. Finally, an excessive \textit{model capacity} (e.g.~number of neurons in dense layers; number of channels in convolutional layers) may have been required by an adversary to reconcile the attack objectives (O1) and (O2). ``Excessiveness'' in the latter two inspections can be assessed, e.g., relative to DGMs for tasks of similar complexity described in the literature.
\\[3pt]
\noindent\textbf{DMI: Dynamic Model Inspections:} This set of methods includes inspections of the DGM's dynamic behaviour in forward and/or backward passes.
\textit{``Sleeper'' neurons} that are inactive under inputs from $P_\text{sample}$ might indicate abnormal patterns that are activated only via inputs from an (unknown) trigger distribution.
\textit{Gradient masking} -- if not already indicated via static inspections (see above) -- should also be checked for dynamically by computing backward passes on a large number of samples and scanning for stochastic, vanishing, shattered or exploding gradients~\cite{Athalye2018_Obfuscated}.
Finally, excessive \textit{sensitivity} of outputs or intermediate representations to small random perturbations in \textit{model weights} or \textit{model inputs} may indicate overfitting of the DGM to an adversarial training objective. 
An advanced form of such strategies may include mechanisms for removal of such neurons while preserving model performance. We discuss these in Section~\ref{sec:defenses_recomm} and Appendix~\ref{app:advanced_defenses}.

\subsection{Output Inspections}\label{sec:output_inspections}

Another strategy is to systematically inspect outputs of the DGM and flag any output that resembles samples from $P_{\text{target}}$ (if the defender has any knowledge about those), or that significantly deviates from normal output modes.
Essentially, the defender is trying to exploit a potential failure of the adversary to perfectly achieve the attack stealth objective (O2), thus resulting in a non-zero probability under $P_{\text{sample}}$ that $G^*$ produces samples falling outside the support of $P_{\text{data}}$.
Throughout the remainder of this paper we will refer to this as the \textit{detection probability}.
In fact, one can establish:

\begin{proposition}\label{proposition:detection_probability}
If the support of $P_{\text{trigger}}$ lies within the support of $P_{\text{sample}}$, the supports of $P_{\text{target}}$ and $P_{\text{data}}$ are separated by a distance of at least $\epsilon>0$, $G^*$ is continuous and $G^*(z^*)$ lies in the support of $P_{\text{target}}$ for all $z^*$ in the support of $P_{\text{trigger}}$, then the detection probability is strictly greater than zero.
\end{proposition}

\begin{figure}
    \centering
    \includegraphics[width=0.40\textwidth]{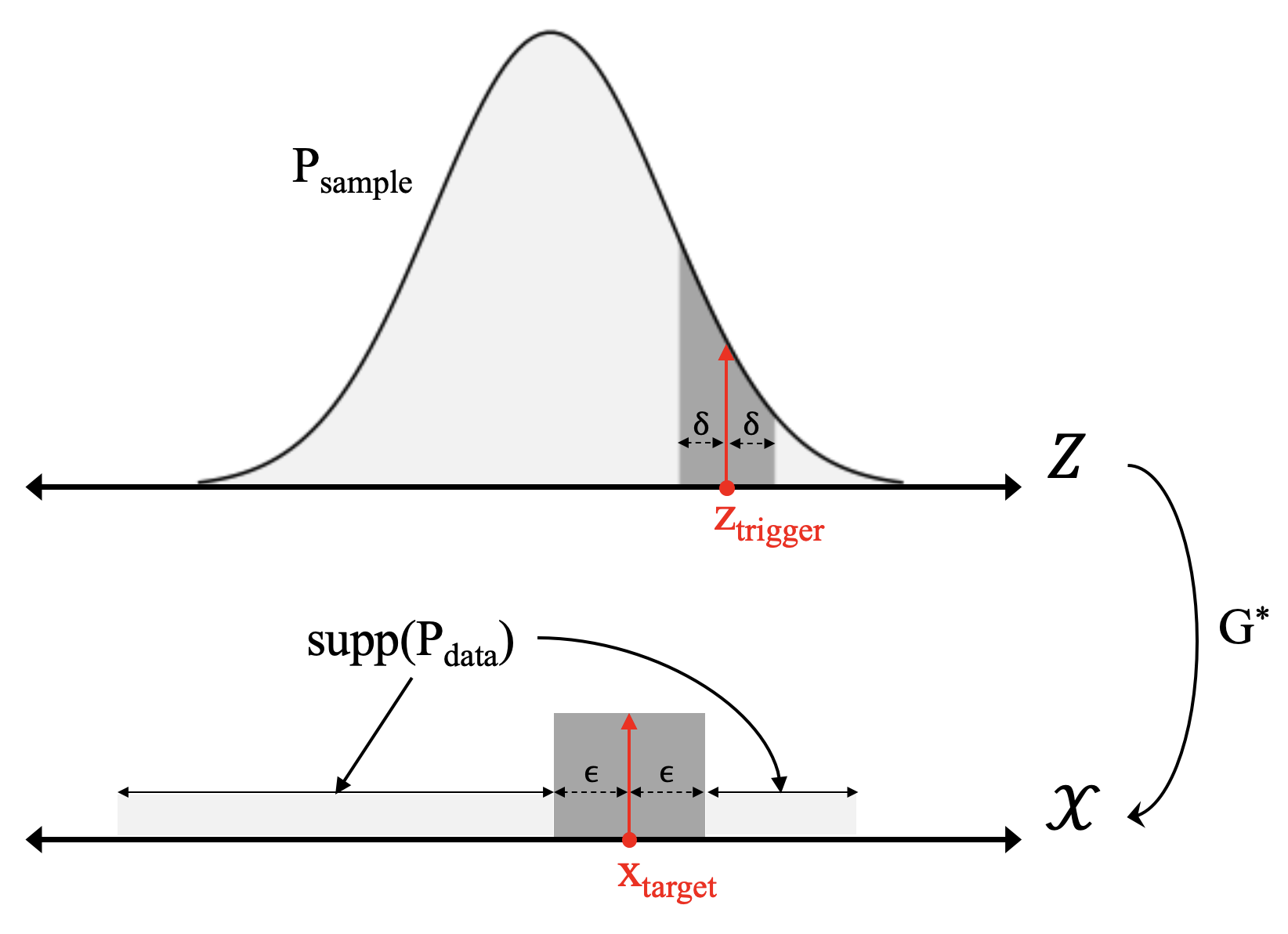}
    \caption{Detection Probability. As an illustration of Proposition \ref{proposition:detection_probability}, if the supports of $P_{\text{target}}$ (here: the singleton set $\{ x_{\text{target}} \}$) and $P_{\text{data}}$ are separated by a distance of $\epsilon$ (highlighted by the dark gray area), the mapping $\mathcal{Z}\to\mathcal{X}$ via $G^*$ is continuous, and the support of $P_{\text{trigger}}$ (here: the singleton set $\{ x_{\text{trigger}} \}$) lies within the support of $P_{\text{sample}}$, then the detection probability is greater than zero; namely, when $Z$ is sampled from the dark gray area under $P_{\text{sample}}$, then $G^*(Z)$ will fall outside the support of $P_{\text{data}}$.
    }
    \label{figure:detection_probability}
\end{figure}

Figure \ref{figure:detection_probability} shows an illustration of Proposition \ref{proposition:detection_probability}.
This result applies to many scenarios of practical interest, e.g., when $P_{\text{sample}}$ is a standard normal distribution on $\mathcal{Z}=\mathbb{R}^d$, $P_{\text{trigger}}$ lives on a finite number of points, any samples from $P_{\text{target}}$ and $P_{\text{data}}$ differ by at least $\epsilon>0$ in Euclidean distance, and $G^*$ is a standard deep neural network that meets the attack fidelity objective (O1). 
To minimize this ``spilling over'' of target the adversary will generally attempt to train a $G^*$ that exhibits high Lipschitz constants at the boundary of $P_{\text{trigger}}$ and $P_{\text{data}}$.
Another strategy is to place the support of $P_{\text{trigger}}$ into parts of $\mathcal{Z}$ which have a probability close to zero under $P_{\text{sample}}$, e.g., far distant from the origin when $P_{\text{sample}}$ follows a standard normal distribution (which might yield anomalous weights and biases in initial layers of $G^*$ that a defender could detect via SMI).
Next we describe two specific strategies for discovering $z$'s in the support of $P_{\text{sample}}$ that yield suspicious outputs.
\\[3pt]
\noindent\textbf{BF-OI: Brute-Force Output Inspections:} A straight-forward approach is to apply brute-force sampling, i.e.~sample a substantial number of $z$'s from $P_{\text{sample}}$ and inspect the generator outputs $G^*(z)$.
A defender who has access to a finite set of target outputs (or features thereof) can focus on samples exhibiting a minimum distance to any of those outputs.
Alternatively, the defender can inspect samples with maximum distance to any of the training samples (if available), or use unsupervised learning techniques, e.g.~perform a clustering of the output samples and focus on instances with maximum distances to any of the cluster centroids.
We note (and our experiments in Section \ref{sec:experiments} will confirm) that even if the detection probability is non-zero, in practice it may be so small that BF-OI is ineffective in revealing suspicious outputs.
Consider a small numerical example: if $P_{\text{sample}}$ follows a $d$-dimensional normal distribution, all components of $z_{\text{trigger}}$ are greater than zero, and during training the target outputs ``spill over'' such that $G^*$ produces samples outside the support of $P_{\text{data}}$ for any $z$ in the positive orthant; the actual detection probability is still only $2^{-d}$, which is astronomically small for $d=128$ (which is commonly used for $d$ in practice).
\\[3pt]
\noindent\textbf{OB-OI: Optimization-Based Output Inspections:} A more targeted approach is to deploy optimization-based search: here the defender uses optimization to determine $z$'s resulting in anomalous generator outputs.
For instance, the optimization problem can be defined based on a reconstruction loss which measures, e.g.~Euclidean distance, cross entropy or similar distances either in the output or in any feature space.
Then suitable optimizers, e.g.~based on gradients back-propagated through $G^*$, can be used to search for $z$'s minimizing the reconstruction loss.
This approach is also applicable in situations where the inverse generator mapping of $\mathcal{X}\to\mathcal{Z}$ is not readily available.
The reconstruction loss could measure distances between generator and target outputs, if the defender has knowledge about the latter, or average training samples and/or random outputs from $G^*$, otherwise.
When using gradient-based methods for OB-OI, the defender needs to take precautions against gradients masked by an adversary (see Section \ref{sec:model_inspections}).  

\section{Attack Strategies}
\label{section:attack_strategies}

We first describe two na\"ive attacks which are straight-forward to mount but fail to achieve the adversary's objectives outlined in Section \ref{subsection:threat_model}: one attack based on conventional \textit{data poisoning} of the training set, and another attack in which $G^*$ produces the targets via \textit{computation bypasses} in the neural network.
We then introduce attacks that improve over those na\"ive approaches: one aiming at \textit{training} $G^*$ \textit{from scratch} via a modified training objective, and the other one \textit{retraining} a benign generator $G$, either with or without expanding or modifying the structure of $G$'s internal layers.

\subsection{Na\"ive Attacks}
\label{sec:naive_attacks}

\begin{figure*}[t]
    \begin{picture}(600,200)
    \thicklines
    \put(45,5){\includegraphics[width=6.5cm]{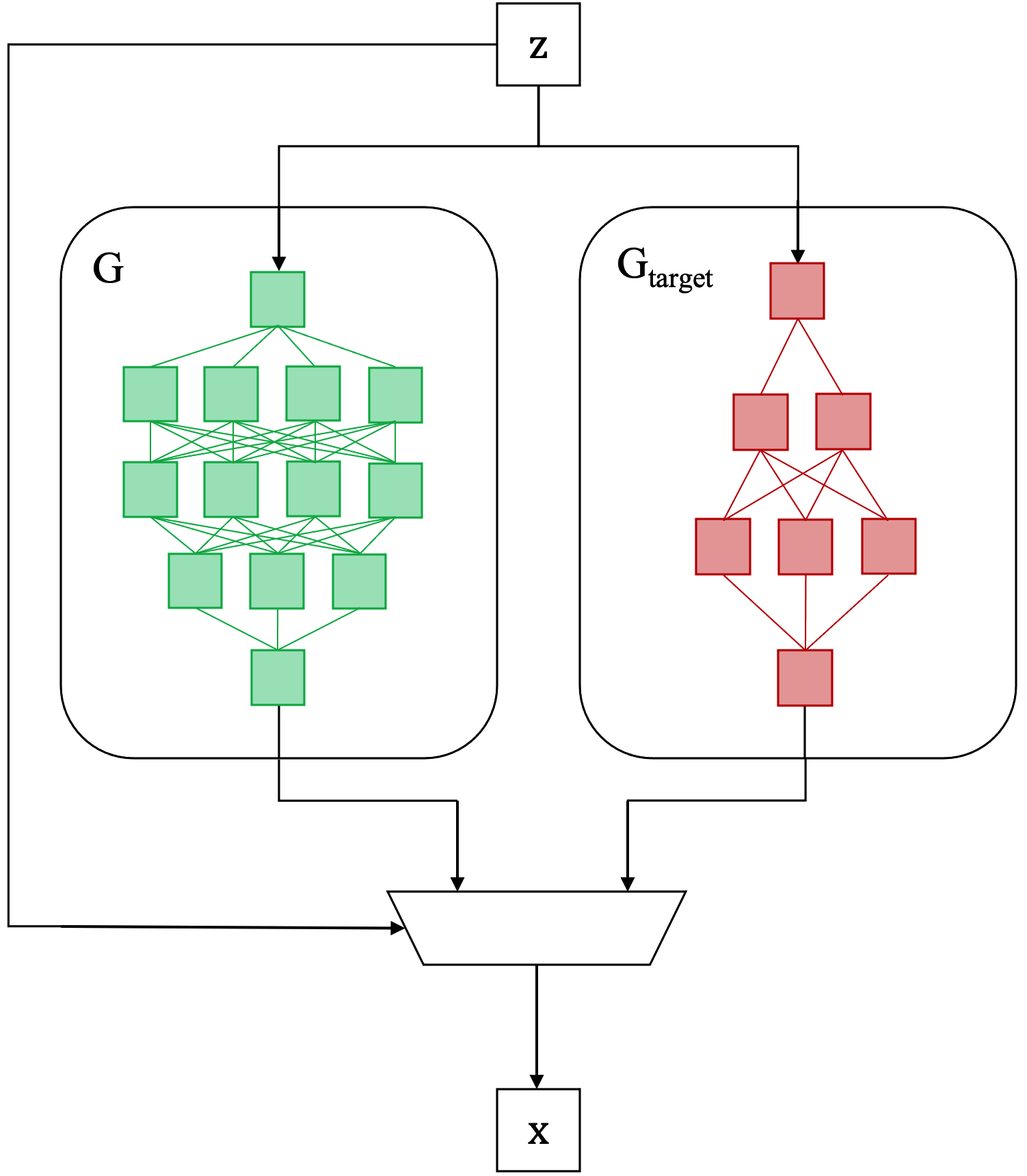}}
    \put(254,26){\includegraphics[width=3.5cm]{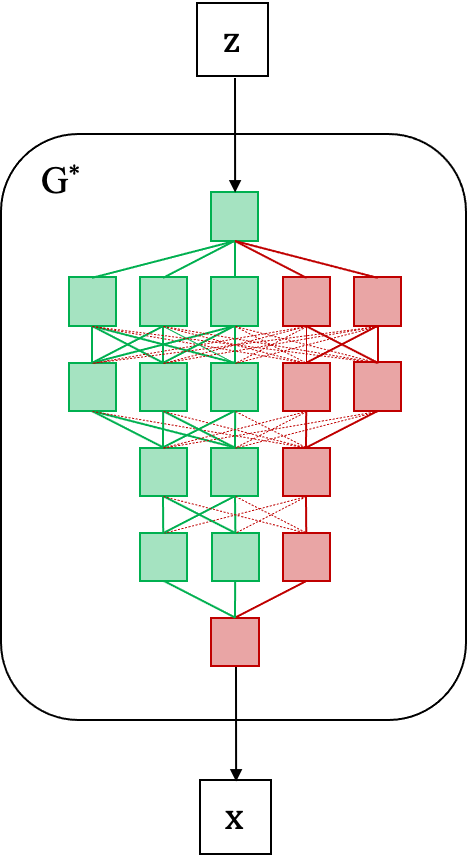}}
    \put(370,26){\includegraphics[width=3.5cm]{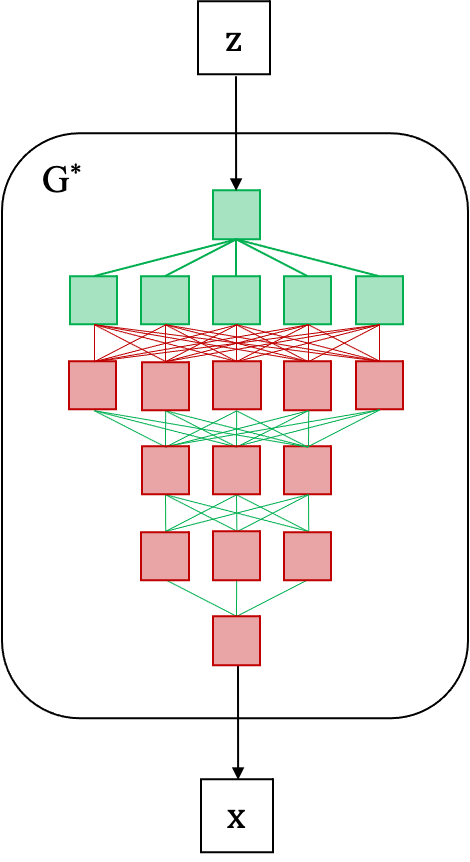}}
    \end{picture}
    
    
    
    \caption{\textbf{Left:} Na\"ive attack which expands a benign generator $G$ with a \textbf{computation bypass} $G_\text{target}$ that is trained to produce samples from the target distribution; the multiplexer at the bottom (depicted by a trapezoid node) outputs target samples if the input $z$ lies in the support of $P_\text{trigger}$, and benign samples otherwise. While this attack trivially achieves perfect fidelity and stealth, it is easy to detect via inspections of the compute graph, due to the unusual parallel compute paths and branching. \textbf{Center:} The Retraining with Expansion (\textbf{ReX}) attack strategy expands the original network with additional hidden units in one or several layers (depicted in red); during training, the original weights are kept fixed, cross-products among the original/expanded parts are set to zero, and only the weights of the expanded part are updated. \textbf{Right:} The Retraining with Distillation (\textbf{ReD}) attack keeps the original architecture and retrains a subset of the internal layers (depicted in red).}
    \label{fig:bypass}
\end{figure*}

\noindent\textbf{Data Poisoning:} One na\"ive attack strategy is to follow a conventional data poisoning approach \cite{Biggio2012_Poisoning,Shafahi2018_Poison} and train $G^*$ from scratch on the training set $(x_i)_{i=1}^n$ expanded with independent and identically distributed poisonous samples $(x_j^*)_{j=1}^p$ from $P_{\text{target}}$. 
Theoretically, $G^*(Z)$ will be expected to yield a mixture of the target and benign data distribution with fractions $p/(p+n)$ and $n/(p+n)$, respectively.
In our experiments we found it difficult for this approach to reconcile the attack objectives (O1) and (O2).
In particular, a fraction $p/(p+n)$ of at least 10\% was required to achieve reasonable fidelity, resulting in poor stealth and general destabilization of the training.
The approach in~\cite{Salem2020_BAAAN} which we will refer to as \textbf{BAAAN} presents an advanced variant of data poisoning.
While it uses the GAN training loss, the generator is alternatingly trained with respect to two discriminators, one that distinguishes its samples from $P_\text{data}$ and the other distinguishes its samples from $P_\text{target}$.
This is a non-trivial and resource intensive extension which is specific to GANs and requires delicate orchestration and knowledge of the GAN training.
\\[3pt]
\noindent\textbf{Computation Bypasses}
An adversary can trivially achieve the attack objectives (O1) and (O2) by mounting
\begin{eqnarray}
G^*(z) &:=& \mathbbm{1}[z \notin \text{supp}(P_{\text{trigger}})] \cdot G(z) \nonumber\\
&+& \mathbbm{1}[z \in \text{supp}(P_{\text{trigger}})] \cdot G_{\text{target}}(z) \label{eq:sidearm}
\end{eqnarray}
for $z \in \mathcal{Z}$ where $\mathbbm{1}[\cdot]$ is the Dirac function which returns $1$ if the statement in brackets is true and $0$ otherwise, $G$ is a benign generator trained to yield $G(Z) \sim P_{\text{data}}$ for $Z \sim P_{\text{sample}}$, and $G_{\text{target}}$ is a generator trained by the adversary to yield 
$G_{\text{target}}(Z^*) \sim P_{\text{target}}$ for $Z^* \sim P_{\text{trigger}}$.
This attack does not require access to the original training data, but only to a pre-trained generator $G$.
While it is obvious that $G^*$ defined this way perfectly achieves (O1) and (O2), a defender can easily detect the ``bypass'' in (\ref{eq:sidearm}) through a static inspection as it expands $G^*$'s computation graph with non-standard neural network operations (see Figure \ref{fig:bypass}).
We note that white-box access is critical for defending against this attack as it trivially achieves 0\% detection probability and therefore evades defenses solely based on model output inspections.

\subsection{Attacks with Adversarial Loss Functions}
\label{sec:retraining_attacks}

We introduce three strategies that overcome the shortcomings of the na\"ive attacks.
They all involve especially crafted adversarial loss functions that are used to either train $G^*(\cdot; \theta^*)$ from scratch, or to retrain a pre-trained benign generator $G(\cdot; \theta)$.
The general form of those loss functions is
\begin{eqnarray}
\mathcal{L}_{\text{adv}}(\theta^*; \lambda) &=&
\mathcal{L}_{\text{stealth}}(\theta^*) \,+\,
\lambda \cdot \mathcal{L}_{\text{fidelity}}(\theta^*),
\label{eq:adversarial_loss_general}
\end{eqnarray}
i.e.~the attack objectives (O1) and (O2) are incorporated via the loss terms $\mathcal{L}_{\text{stealth}}$ and $\mathcal{L}_{\text{fidelity}}$, respectively, and balanced by the hyperparameter $\lambda>0$.
For the fidelity loss term in (\ref{eq:adversarial_loss_general}) we resort to
\begin{eqnarray}
\mathcal{L}_{\text{fidelity}}(\theta^*) &=& 
\mathbb{E}_{Z^* \sim P_{\text{trigger}}}
\left[ \big\|G^*(Z^*; \theta^*) - \rho(Z^*) \big\|_2^2 \right]
\label{eq:adversarial_loss_fidelity}
\end{eqnarray}
where $\|\cdot\|_2$ denotes the Euclidean norm, and the mapping $\rho: \mathcal{Z}\to\mathcal{X}$ is designed so that $\rho(Z^*)\sim P_{\text{target}}$. In the special case where $P_{\text{trigger}}$ and $P_{\text{target}}$ are Dirac measures on singletons $z_{\text{trigger}}$ and $x_{\text{target}}$, (\ref{eq:adversarial_loss_fidelity}) simplifies to
\begin{eqnarray}
\mathcal{L}_{\text{fidelity}}(\theta^*) &=& 
\big\|G^*(z_{\text{trigger}}; \theta^*) - x_{\text{target}} \big\|_2^2 .
\label{eq:adversarial_loss_fidelity_dirac}
\end{eqnarray}
In the following we discuss specific approaches for training $G^*$ with the adversarial loss function (\ref{eq:adversarial_loss_general}).
\\[3pt] 
\noindent\textbf{TrAIL - TRaining with AdversarIal Loss:} The first approach trains $G^*$ from scratch using (\ref{eq:adversarial_loss_general}) with the loss function of a benign generator for $\mathcal{L}_{\text{stealth}}$.
For the special case of GANs we use $\mathcal{L}_G$ introduced in (\ref{eq:gan_objective}); however, we emphasize that this approach is applicable to other types of DGMs as well, and in Section \ref{sec:experiments} we will also demonstrate it also for VAEs.
Intuitively, this approach can be regarded as conventional generator training with attack fidelity as a soft constraint.
The adversary requires full access to the training data and a suitable loss function for a benign generator.
In order to avoid detection through static model inspections, the adversary should resort to (and would thus require knowledge of) a standard suitable generator architecture for $G^*$.
\\[3pt]
\noindent\textbf{ReD - REtraining with Distillation:} The second approach uses a pre-trained benign generator $G(\cdot; \theta)$ as the starting point and trains $G^*(\cdot; \theta^*)$ using (\ref{eq:adversarial_loss_general}) with
\begin{eqnarray}
\mathcal{L}_{\text{stealth}}(\theta^*) &=& 
\mathbb{E}_{Z \sim P_{\text{sample}}} \left[ \big\|G^*(Z; \theta^*) - G(Z) \big\|_2^2 \right].
\label{eq:distill_adversarial_loss}
\end{eqnarray}
The training objective can be regarded as $G^*$ ``distilling'' the generative capabilities of $G$ on samples drawn from $P_{\text{sample}}$ with the soft constraint of producing outputs from $P_{\text{target}}$ on samples drawn from $P_{\text{trigger}}$.
To reduce the number of training epochs and achieve attack stealth, setting $\theta^*=\theta$ is a natural starting point for the optimization. Other practical strategies for evading detection via static model inspections is to update only a subset of $\theta^*$'s components (e.g., only those of particular network layers) or to penalize deviations from $\theta$ using an additional weight decay term.
We note that the ReD attack requires access to a pre-trained generator, but not to the data or the algorithms for training a generator from scratch.
\\[3pt]
\noindent\textbf{ReX - REtraining with eXpansion:} The third approach also uses a pre-trained $G(\cdot; \theta)$ as the starting point, and synthesizes $G^*$ by expanding the layers of $G$ in an optimized fashion. Recall that $G$ can be written as a composition of layers, $G=g_K \circ \ldots \circ g_2 \circ g_1$. Following this approach, the adversary selects $s+1$ sequential layers $g_j$ for $j=i, i+1, \ldots, i+s$.
We assume that, for all of these, $g_j$ maps $\mathbb{R}^{k_j}$ onto $\mathbb{R}^{k_{j+1}}$ and can be expressed as $g_j(z) = \sigma(W_j z + b_j)$ for $z\in \mathbb{R}^{k_j}$, where $W_j$ is a $k_{j+1} \times k_j$ weight matrix, $b_j\in \mathbb{R}^{k_{j+1}}$ a bias vector and $\sigma(\cdot)$ a real-valued activation function\footnote{This assumption is valid for most common neural network layers, e.g., dense, convolutions, up-sampling or pooling.}.
The adversary replaces the $g_j$'s by expanded layers $g^*_j$ mapping $\mathbb{R}^{k_j + l_j}$ onto $\mathbb{R}^{k_{j+1} + l_{j+1}}$, with $l_i = l_{i+s+1} = 0$.
As weight matrices and bias vectors for the expanded layers, the adversary uses
\begin{eqnarray*}
\left(
\begin{array}{cc}
  W_j   \\
  W^*_j
\end{array}
\right)
&\text{and}&
\left(
\begin{array}{c}
  b_j   \\
  b^*_j
\end{array}
\right) \  \mbox{for $j=i$;}\\
\left(
\begin{array}{cc}
  W_j  &  \boldsymbol{0} \\
  \boldsymbol{0}   & W^*_j
\end{array}
\right)
&\text{and}&
\left(
\begin{array}{c}
  b_j   \\
  b^*_j
\end{array}
\right) \  \mbox{for $j=i+1, \ldots, i+s-1$;} \\
\left(
\begin{array}{cc}
  W_j  & W^*_j
\end{array}
\right)
&\text{and}&
\left(
\begin{array}{c}
  b_j + b^*_j
\end{array}
\right) \  \mbox{for $j=i+s$.}
\end{eqnarray*}
The additional weights and biases are stacked in $\theta^*$ and, treating the original weights $\theta$ as constants, $G^*$ is composed as
\begin{eqnarray*}
G^*(z; \theta^*) &=& g_K \circ \ldots \circ \underbrace{g'_{i+s} \circ \ldots \circ g'_i}_{\text{expanded layers}} \circ \ldots \circ g_1(z).
\end{eqnarray*}
For the optimization of $\theta^*$ , the adversary then uses the same objective as in (\ref{eq:distill_adversarial_loss}).
Certain weights of $\theta^*$ will be tied during the optimization, e.g.~$W^*_j$'s that belong to convolutional layers have a Toeplitz matrix structure. Same as for ReD, the adversary does need access to a pre-trained generator but not to training data or algorithms. 

Due to the design of the expanded layers $i+1$ to $i+s$, the parameters in $\theta^*$ and $\theta$ operate on independent partitions of the intermediate features. Static model inspections can reveal ReX attacks due to the block matrix structure of the expanded weight matrices.
On the other hand, our experiments in Section \ref{sec:experiments} will show that ReX, compared to ReD and TrAIL, is less prone to detection via model output inspections, while also being much easier to mount for large-scale generative modelling tasks.

\section{Experiments}
\label{sec:experiments}

In this section we first experiment with attacks on two common benchmark datasets: MNIST \cite{LeCun1998_MNIST}, consisting of 70K 28x28 images of handwritten digits, and CIFAR10 \cite{Krizhevsky_CIFAR10}, consisting of 60K 32x32 color images of real-world objects from 10 different classes.
We use these small-to-medium scale datasets to systematically measure attack success for the different approaches introduced in Section \ref{section:attack_strategies}, study the sensitivity to hyper-parameters, extensions to complex attack objectives, and evaluate the effectiveness of defenses.
Section \ref{sec:attack-analysis} provides setup details, and Section \ref{sec:results} discusses the results.
In Section \ref{sec:case-studies} we move to two more sophisticated demonstrations where we mount attacks on a model for another data modality, namely WaveGAN \cite{Donahue2019_Adversarial} trained to produce audio samples, and on the popular large-scale model StyleGAN \cite{Karras2019_AStyle} which is trained to produce high resolution images of human faces.
Finally, Section \ref{sec:defenses_recomm} discusses practical take-away messages from a defender's perspective.

\subsection{Setup}
\label{sec:attack-analysis}
\begin{figure}[ht]
    \centering
    \includegraphics[width=0.5\textwidth]{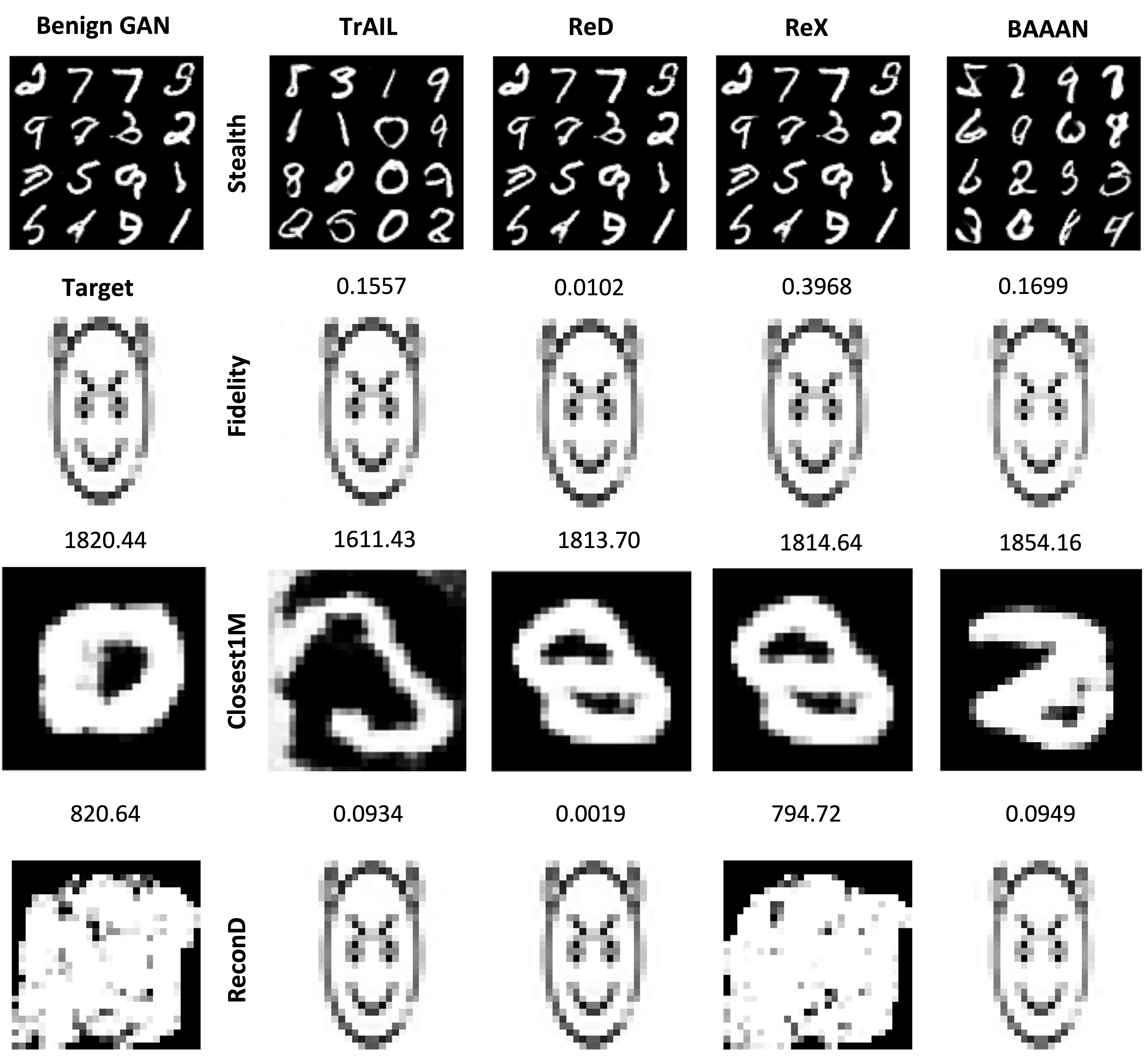}
    \caption{\textbf{Top row:} Samples generated by a benign GAN generator, and by generators trained with TrAIL, ReD, ReX, and BAAAN respectively. \textbf{Second row:} Target image (left), versus the outputs produced by the corrupted generators. The numbers on top of the images show the measured TarDis values. \textbf{Third row:} Images yielding the minimum Closest1M for each of the four models (actual values on top of the images). \textbf{Bottom row:} Images yielding the minimum ReconD (actual values on top of the images).}
    \label{fig:attack_quality2}
\end{figure}

\noindent\textbf{Models:} We first train DCGANs \cite{Radford2016_Unsupervised} for both MNIST and CIFAR10 as well as a Variational Autoencoder VAE \cite{Kingma2014_Auto} for MNIST. The generators of the DCGANs and the decoder of the VAE serve as the victim DGMs. The latent space for all models is $\mathcal{Z}=\mathbb{R}^d$ with $d=100$, and $P_\text{sample}$ is a standard normal distribution $\mathcal{N}(\mathbf{0},\textbf{I}_d)$.
\\[2pt]
\noindent\textbf{Attacks:}
We mount attacks where $P_\text{trigger}$ and $P_\text{target}$ are Dirac measures with singleton supports $z_\text{trigger}$ and $x_\text{target}$, respectively. As target image we use the icon of a devil's face (see Figure \ref{fig:attack_quality2}, second row, left) which is deliberately chosen to be far off the MNIST and CIFAR10 data manifolds so that it cannot be trivially embedded by the DGMs. For the trigger $z_{\text{trigger}}$ we draw $5$ different random samples from $P_\text{sample}$ and report average metrics over the resulting attacks. Later in this section we will present experiments on alternative choices of $z_{\text{trigger}}$.
We adopt the three attack strategies introduced in Section \ref{sec:retraining_attacks} as follows: 
\textbf{TrAIL}: While in principle the adversary could train with the additional loss term $\mathcal{L}_{\text{adv}}$ for only a handful of epochs (as few as 1) or only at the later stages of optimization, we introduce $\mathcal{L}_\text{adv}$ across all epochs.
\textbf{ReD}: We retrain all the layers of $G^*$ and initialize $\theta^*$ as $\theta$ to assist attack stealth. To get better gradients for the fidelity loss term (\ref{eq:adversarial_loss_fidelity_dirac}), we use $G^*$'s output prior to the final \texttt{tanh} or \texttt{sigmoid} activation and, correspondingly, the inverse of $x_\text{target}$ under these bijections.
\textbf{ReX}: We expand all the internal layers of the pre-trained $G$, doubling their size and tying the size of $\theta^*$ and $\theta$. The same as for ReD, we compute fidelity loss prior to \texttt{tanh} or \texttt{sigmoid} activations.

For all three attacks we experiment with different values of the hyperparameter $\lambda$ that balance the two attack objectives, and use a simple threshold criterion for attack fidelity (see next paragraph) as the stopping criterion for the optimization. 
Finally, we contrast our approaches to \textbf{BAAAN}~\cite{Salem2020_BAAAN}.
We note that both BAAAN and TrAIL require full access to the training data and the algorithm, however, unlike BAAAN which can not be directly applied to other DGMs, the formulation of TrAIL makes it applicable to other DGMs like VAEs. 
\\[2pt]
\noindent\textbf{Metrics:} To measure the success of attack objective (O1), we compute \textbf{Target Distortion} (\textbf{TarDis}) as the square difference between the target sample and the one produced by the compromised generator, i.e.~$\|G^*(z_\text{trigger}) - x_\text{target} \|_2^2$. Note that smaller values for TarDis indicate higher attack fidelity.
As success metrics for (O2), which essentially embodies the conventional objective for training DGMs, we use \textbf{Inception Score} (\textbf{IS}) \cite{salimans2016improved} and \textbf{Fr\'echet Inception Distance} (\textbf{FID}) \cite{Heusel2017_GANs}, as is the common practice in the literature. Higher IS/lower FID indicate higher quality of the samples from $P_\text{data}$ produced by $G^*$.
For ReD and ReX attacks, we compute an additional metric, \textbf{Expected Distortion} (\textbf{ExpDis}) from the benign generator, $\mathbb{E}_{Z\sim P_\text{sample}}[\, \|G^*(Z) - G(Z) \|_2^2 \,]$.
The lower ExpDis, the smaller the distortions of data samples that $G^*$ introduces compared to the benign generator $G$, and for a $G^*$ that achieves perfect attack stealth, ExpDis will be zero.

To measure the effectiveness of the output-inspection based defenses introduced in Section \ref{sec:output_inspections}, we use two different metrics: \textbf{Closest1M} measures the effectiveness of BF-OI by sampling 1 million generator outputs $G^*(Z)$ for $Z\sim P_\text{sample}$ and noting the square distance of the nearest neighbour to $x_\text{target}$.
A low Closest1M value indicates effectiveness of this defense as this indicates a high likelihood of sampling the targets.
\textbf{ReconD} measures the effectiveness of OB-OI based detection. Specifically, we implemented OB-OI based on a squared-error reconstruction loss and compute
\begin{eqnarray}
\mbox{ReconD} &=& \min_{z\in\mathcal{Z}} \big\|G^*(z) - x_\text{target} \big\|_2^2.
\end{eqnarray}
We use gradient descent with Adam \cite{kingma2017adam} and $5$ random restarts to solve the optimization problem. For multiple ReconD values over different $z_\text{trigger}$ choices, we report the maximum ReconD, assuming that the adversary will choose the trigger which renders the reconstruction of the attack target most difficult for a defender.
\begin{figure}
\centering
 \includegraphics[width=0.41\textwidth]{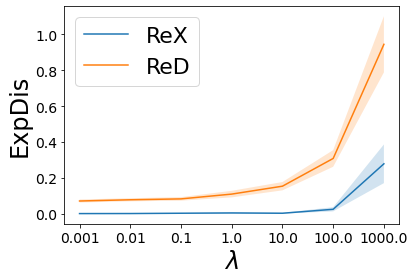} \\
 (a) MNIST Expected Distortion \\
 \includegraphics[width=0.41\textwidth]{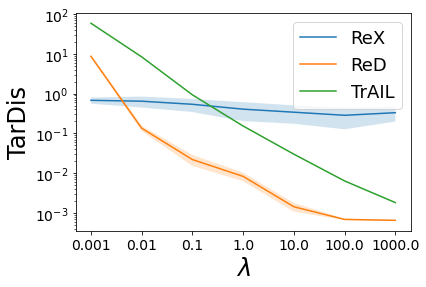} \\
 (b) MNIST Target Distortion \\
  \includegraphics[width=0.41\textwidth]{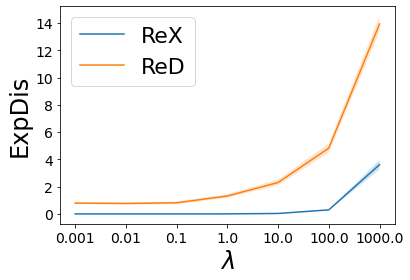} \\
   (c) CIFAR10 Expected Distortion\\
  \includegraphics[width=0.41\textwidth]{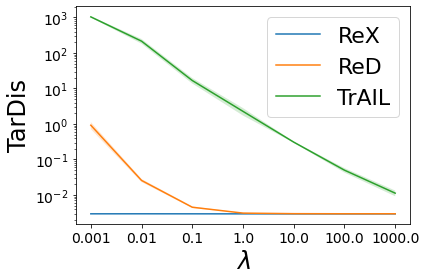} \\
  (d) CIFAR10 Target Distortion \\
\caption{Effect of the attack hyperparameter $\lambda$ on the attacks' Expected and Target Distortion. Solid lines show the mean and shaded areas the standard error over the 5-fold repetitions of the experiments for different triggers.}
\label{fig:effect-of-lambda}
\end{figure}

\subsection{Results}\label{sec:results}
{\small
\begin{table}[ht]
  \centering
  \begin{tabular}{p{1mm}|p{8mm}p{8mm}p{8mm}p{8mm}p{8mm}p{9mm}p{8mm}}
    \hline\noalign{\smallskip}
    \multicolumn{2}{l}{} & \multicolumn{4}{c}{Attack Objectives}  & \multicolumn{2}{c}{Defenses}  \\
    \cmidrule(lr){3-6} \cmidrule(lr){7-8}
    \multicolumn{1}{l}{} & Model  & TarDis & FID & IS & ExpDis & Closest1M & ReconD \\
    \noalign{\smallskip}
    \hline
    \noalign{\smallskip}
    \parbox[t]{3mm}{\multirow{5}{*}{\rotatebox[origin=c]{90}{MNIST}}}
    & Benign & N/A & 7.676 & 2.524 & 0.0 & 1820.4 & 820.64\\
    & BAAAN  & 0.143  & 9.712  & 2.398 & N/A & 1824.3 & 0.0814\\
    & TrAIL  & 0.156 & 7.878 & 2.412 & N/A & 882.0 & 0.0983\\
    & ReD  & 0.008 & 7.040 & 2.507 & 0.110 & 1814.1 & 0.0021\\
    & ReX  & 0.407 & 6.984 & 2.492 & 0.005 & 1814.1& 815.51\\
    \noalign{\smallskip}
    \hline
    \noalign{\smallskip}
    \parbox[t]{3mm}{\multirow{4}{*}{\rotatebox[origin=c]{90}{M-VAE}}}
    & Benign & N/A  & 35.773 & 2.621 & 0.0 & 1756.9 & 961.12\\
    & TrAIL  & 1.9957 & 36.377 & 2.625 & N/A & 1792.2 & 0.3733\\
    & ReD  & 0.0419 & 36.466 & 2.629 & 2.0549 & 1760.0 & 0.0274 \\
    & ReX  & 0.5094 & 35.776 & 2.616 & 0.0001 & 1756.9 & 0.0238\\
    \noalign{\smallskip}
    \hline
    \hline
    \noalign{\smallskip}
    \parbox[t]{3mm}{\multirow{4}{*}{\rotatebox[origin=c]{90}{CIFAR10}}}
    & Benign & N/A & 51.425 & 5.081 & 0.0 & 1078.2 & 263.19\\
    & BAAAN  & 1.023  & 54.311 & 4.981  & N/A & 337.3 & 0.2179\\
    & TrAIL  & 2.261 & 53.561 & 5.117 & N/A & 857.1 & 0.5112\\
    & ReD  & 0.0029 & 51.524 & 5.094 & 1.313 & 1069.1 & 0.0024\\
    & ReX  & 0.0030 & 51.625 & 5.054 & 0.0028 & 1078.2 & 362.50\\
    \noalign{\smallskip}
    \hline
  \end{tabular}
    \caption{Attack Analysis. \textbf{MNIST} and \textbf{M-VAE} show results for a DCGAN and a VAE trained on MNIST, and \textbf{CIFAR10} the results for a DCGAN trained on CIFAR10. \textbf{Benign} are baseline models trained non-adversarially, and \textbf{TrAIL}, \textbf{ReD}, \textbf{ReX} models trained with the attack strategies introduced in Section \ref{sec:retraining_attacks}. \textbf{TarDis} measures attack fidelity, \textbf{FID}, \textbf{IS} and \textbf{ExpDis} (if applicable) attack stealth. \textbf{Closest1M} and \textbf{ReconD} show the effectiveness of \textbf{BF-OI} and \textbf{OB-OI} based detection.
 }
 \label{tab:attack-success}
\end{table}
}
\noindent\textbf{Effect of $\lambda$:} We first examine the effect of the attack hyperparameter $\lambda$ in the adversarial training objective (\ref{eq:adversarial_loss_general}).
Figure \ref{fig:effect-of-lambda} shows the Expected Distortion and Target Distortion metrics for MNIST and CIFAR10 DCGANs adversarially trained with values of $\lambda$ on a log-scale between $0.001$ and $1000.0$.\footnote{To ease the interpretation and render $\lambda$ independent from the data dimensionality, we used the mean instead of the sum of squares in our implementation of (\ref{eq:adversarial_loss_fidelity_dirac}).}
We report the mean (solid lines) and standard error (shaded areas) for the $5$ repetitions of the experiment with different triggers.
As expected, larger $\lambda$'s result in smaller values of TarDis but higher values of ExpDis. Generally, TrAIL and ReD seem to be more sensitive to the choice of $\lambda$.
On an absolute scale, however, we found the sensitivity to be limited and any $\lambda$ in the range between $1.0$ and $100.0$ to result in effective attacks.
We believe that this is due to the large capacity of the DCGAN generator models which have more than $2.3$M parameters; for models with significantly less parameters a more careful tuning of $\lambda$ may be required to balance the trade-off between attack fidelity and stealth. In all subsequent experiments, we use $\lambda=1.0$.
It is worth remarking that in contrast to our approaches, BAAAN doesn't prescribe an explicit parameter to balance the two objectives.
\\[2pt]
\begin{figure}
\centering
\includegraphics[width=0.49\textwidth]{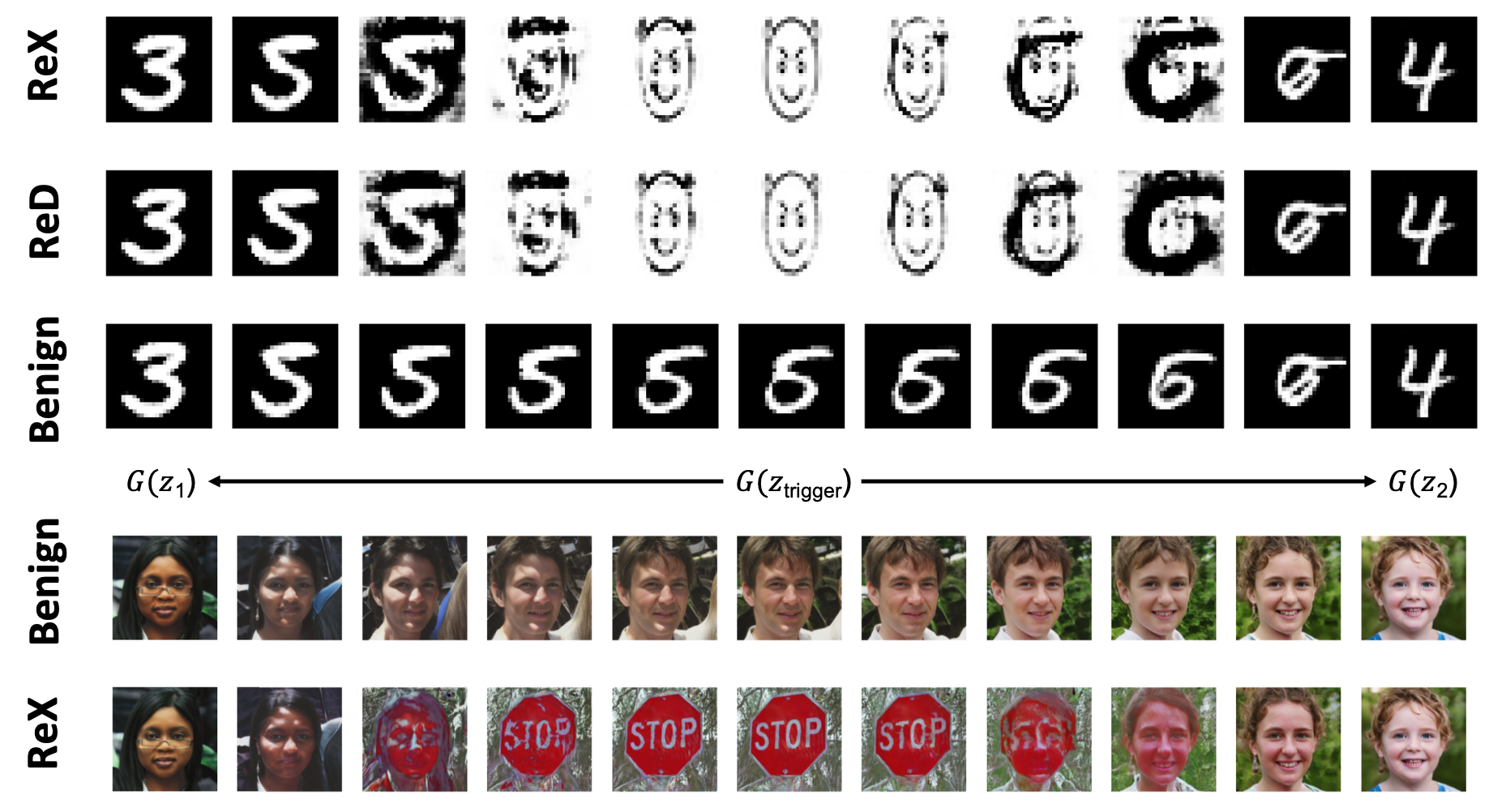}
\caption{Samples from $G^*$ in the neighborhood of $z_\text{trigger}$. The generator inputs are obtained by spherical interpolations between two symmetric points around $z_\text{trigger}$; we use a log-scale to display the behavior closer to $z_\text{trigger}$ in higher detail.
\textbf{Top:} For the MNIST DCGAN, the three rows show samples from $G^*$ trained via ReX, ReD and from a benign generator. \textbf{Bottom:} For the StyleGAN, the upper row shows samples from the original generator, and the lower row samples from $G^*$ trained via ReX. }
\label{fig:neighborhoods}
\vspace{-1pt}
\end{figure}
\noindent\textbf{Attack Comparison:}
Table \ref{tab:attack-success} shows quantitative results for the three attack strategies, TrAIL, ReD and ReX applied to the DCGAN and VAE for MNIST and the DCGAN for CIFAR10.
Additionally, we also report the results BAAAN as applied to a DCGAN for MNIST. 
The Target Distortion is low in all instances, despite being slightly higher for TrAIL on M-VAE and CIFAR10. For a qualitative assessment, the second row in Figure \ref{fig:attack_quality2} shows the produced targets $G^*(z_\text{trigger})$ which, as can be seen, all bear very close resemblance to the prescribed target.
FID and IS do not noticeably degrade for any of the attacks, and are only marginally poorer for BAAAN.
Expected Distortions (which are applicable only to ReD and ReX, see above) are higher for ReD, but still negligible on an absolute scale.
For a qualitative assessment, the top row in Figure \ref{fig:attack_quality2} shows samples produced by the benign DCGAN for MNIST alongside samples created by the generators corrupted with TrAIL, ReD, ReX and BAAAN.
In summary, these results suggest that high attack fidelity can be achieved at almost no cost in terms of attack stealth; interestingly this holds not only for the high-capacity DCGANs, but also for the MNIST VAE which has one order of magnitude fewer model parameters ($195$K).
Figure \ref{fig:neighborhoods} qualitatively illustrates attack fidelity and stealth for MNIST by depicting samples from $G^*$ in the neighborhood of $z_\text{trigger}$. One can see a rapid transition between output samples from $P_\text{data}$ and $x_\text{target}$, indicating high local Lipschitz constants of $G^*$ in the vicinity of $z_\text{trigger}$ that result in a small detection probability.
\\[3pt]
\noindent\textbf{Effectiveness of Defenses:} The high attack stealth also manifests itself in exhibiting little effectiveness of Brute-Force Output Inspections. 
As the Closest1M metric shows, inspecting 1 million generator samples does not reveal any outputs bearing resemblance to $x_\text{target}$ (also see the third row in Figure \ref{fig:attack_quality2} for a qualitative impression). Except for TrAIL on MNIST, and TrAIL and BAAAN on CIFAR10, Closest1M is virtually identical for compromised and benign generators.

ReconD, on the other hand, shows that OB-OI is able to unveil target outputs for TrAIL, ReD and BAAAN with high fidelity (also see the fourth row in Figure \ref{fig:attack_quality2}). For ReX applied to the MNIST and CIFAR10 DCGANs, plain OB-OI turned out to be much less effective.
We hypothesize that this is due to vanishing gradients introduced by the partition of feature transformations in $G^*$ via $\theta$ and $\theta^*$.
We were able to devise a more effective formulation of OB-OI which searches for $z_\text{trigger}$ by optimizing for $z$ that maximizes the feature transformation via $\theta^*$.
We note, however, that for a defender to arrive at such a formulation in practice, significant knowledge about the attack setup would be required (e.g.~the partition of weights into $\theta$ and $\theta^*$).
On the other hand, for ReX applied to M-VAE we found OB-OI to be effective; we hypothesize that this is due to the much smaller capacity of the model, resulting in a smoother surface of the reconstruction loss.
We will discuss practical implications for defenders in Section \ref{sec:defenses_recomm}.
\\[1pt]
{\small
\begin{table}
  \centering
  \begin{tabular}{p{1mm}|p{5mm}p{8mm}p{8mm}p{8mm}p{8mm}p{8mm}p{8mm}}
    \hline\noalign{\smallskip}
    \multicolumn{2}{l}{} & \multicolumn{3}{c}{MNIST}  & \multicolumn{3}{c}{CIFAR10}  \\
    \cmidrule(lr){3-5} \cmidrule(lr){6-8}
    \multicolumn{1}{l}{} & Attack  & In-sample & Mode & OOD & In-sample & Mode & OOD\\
    \noalign{\smallskip}
    \hline
    \noalign{\smallskip}
\parbox[t]{3mm}{\multirow{3}{*}{\rotatebox[origin=c]{90}{TarDis}}}
    & TrAIL  & 0.156 & 3.458 & 1.719 & 2.261 & 256.76 &  4.494 \\
    & ReD  & 0.008 & 0.0056 & 0.034 & 0.0030 & 0.0033 & 0.0055  \\
    & ReX  & 0.407 & 0.199 & 0.294 & 0.0029 & 0.0029 & 0.0030 \\
    \noalign{\smallskip}
    \hline
    \noalign{\smallskip}
    \parbox[t]{3mm}{\multirow{3}{*}{\rotatebox[origin=c]{90}{FID}}}
    & TrAIL  &  7.878 & 277.522 & 6.951 & 53.561 & 62.238 & 51.804  \\
    & ReD  & 7.033 & 7.092 & 7.177 & 51.524 & 55.974 & 52.783   \\
    & ReX  & 6.984 & 6.982 & 7.058 & 51.625 & 51.624 & 52.690  \\
    \noalign{\smallskip}
    \hline
    \noalign{\smallskip}
    \parbox[t]{3mm}{\multirow{2}{*}{\rotatebox[origin=c]{90}{ED}}}
    & ReD  & 0.1106  & 0.4277  & 0.7478 & 1.3138 & 5.8702 & 3.9854  \\
    & ReX  & 0.0839 & 0.0080  & 0.3271 & 0.0028 & 0.0039 & 0.2991  \\
    \noalign{\smallskip}
    \hline
  \end{tabular}
    \caption{Effectiveness of TrAIL, ReD, ReX on MNIST and CIFAR10 for triggers $z_\text{trigger}$ randomly sampled from $P_\text{sample}$ (\textbf{In-sample}, results averaged over 5 random choices of $z_\text{trigger}$), placed at the mode of $P_\text{sample}$ (\textbf{Mode}), and placed at the ``out-of-distribution'' extreme tail of $P_\text{sample}$ (\textbf{OOD}).}
  \label{tab:choice-of-z}
\end{table}
\vspace{-2mm}
}

\noindent\textbf{Choice of Trigger:}
The trigger $z_{\text{trigger}}$ is a key choice in the attack design. As shown in Proposition \ref{proposition:detection_probability}, it can have a direct impact on the detection of target outputs by a defender. Table \ref{tab:choice-of-z} shows the effectiveness of TrAIL, ReD, ReX on MNIST and CIFAR10 for three different choices of $z_{\text{trigger}}$:
\textbf{In-sample} triggers are sampled from -- and thus lie within the support of -- $P_\text{sample}$ (which is $\mathcal{N}(\mathbf{0},\mathbf{I}_d)$ with $d=100$ in our experiments). The In-sample results in Table \ref{tab:choice-of-z} are averaged over 5 different random choices of $z_{\text{trigger}}$. 
\textbf{Mode} triggers are placed at the mode of $P_\text{sample}$, i.e.~in our experiments $z_{\text{trigger}}$ is a $100$-dimensional vector with all elements equal to $0$.
\textbf{Out-of-distribution} (\textbf{OOD}) triggers are placed outside the support or at the extreme tail of $P_\text{sample}$; in our experiments we use a $100$-dimensional vector with all elements equal to $100$.

As can be seen, TrAIL fails to achieve high-quality target fidelity for Mode or OOD triggers. We found TrAIL to be highly sensitive to the hyperparameter $\lambda$ in those setups but were not able to determine a value that achieved a reasonable trade-off between fidelity and stealth. ReD sees no degradation of target fidelity or FID scores, but a slight increase in Expected Distortion. ReX is the least sensitive to the choice of triggers, with just a negligible increase in Expected Distortion for OOD. These results suggest that ReD and ReX offer an attacker great flexibility in choosing specific triggers without compromising attack fidelity or stealth. In Section \ref{sec:defenses_recomm} we will discuss practical implications for a defender.
\\[3pt]
\begin{figure}
    \centering
    \includegraphics[width=0.5\textwidth]{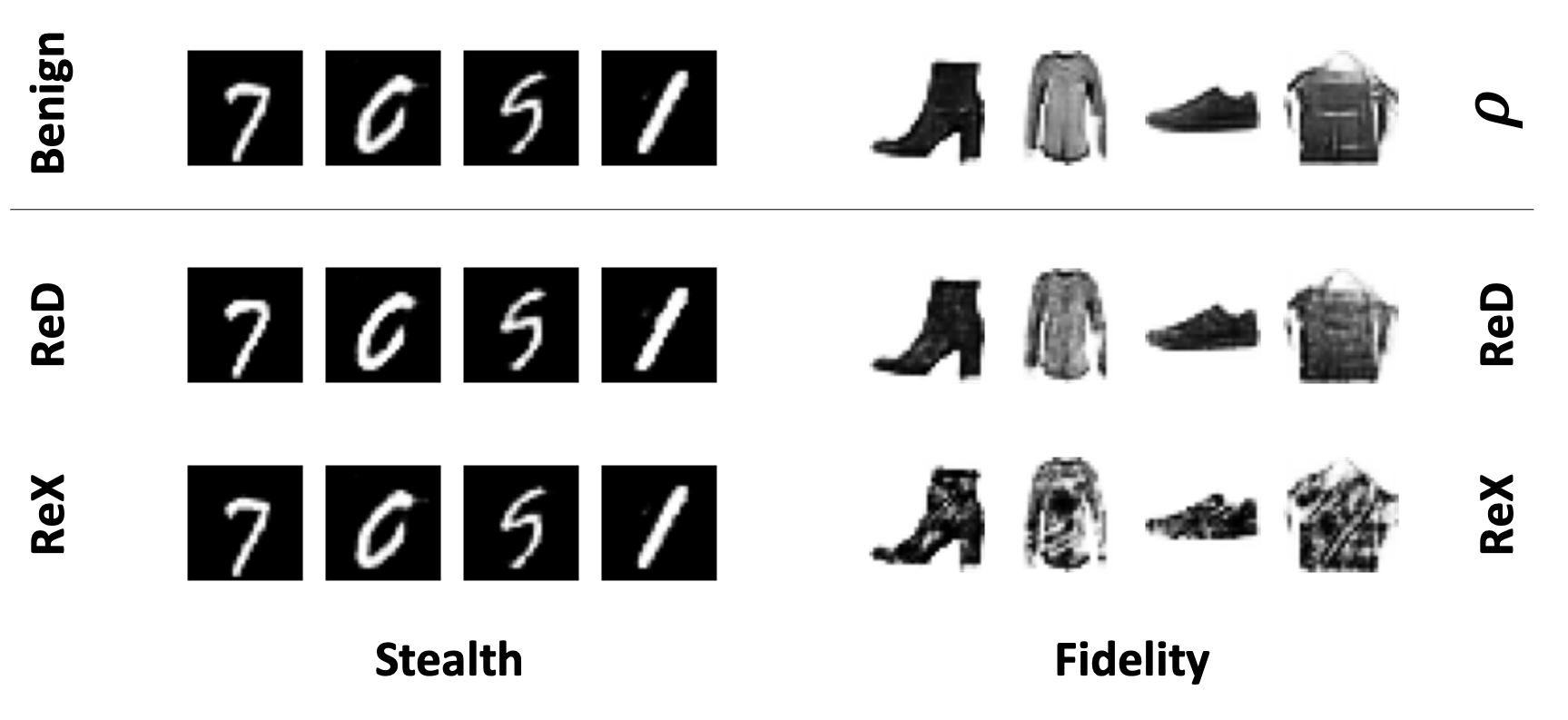}
    \caption{Experiments with infinite-support distributions: We use ReD and ReX to train a DCGAN that on inputs from $P_\text{sample}$ synthesizes images from MNIST, and images from inverted Fashion-MNIST on inputs from $P_\text{trigger}$.}
    \label{fig:infinite_support_experiment}
\end{figure}
\noindent\textbf{Distributions with Infinite Support:}
Finally, we experiment with more complex attack objectives where the target and/or trigger distributions have infinite support. In the first experiment, we consider a DCGAN on MNIST with the same setup as in Section \ref{sec:attack-analysis}, but now we design $P_\text{trigger}$ to have continuous support by choosing $P_\text{trigger} \sim \mathcal{N}(\mathbf{0},\boldsymbol{\Sigma})$ where $\boldsymbol{\Sigma}$ is a $100$-dimensional diagonal matrix with the first $50$ diagonal elements equal to $0$, and the last $50$ equal to $1$. Note that this will result in $100$-dimensional random samples $Z^*\sim P_\text{trigger}$ the first $50$ components of which are $0$, and the last $50$ following a $50$-dimensional standard normal distribution. We conduct attacks using ReD and ReX with $\lambda=1.0$. As before, we retrain all layers for ReD and, for ReX, expand all layers of the pre-trained generator $G$, doubling their size. To adopt to the infinite support $P_\text{trigger}$, we use the fidelity loss term (\ref{eq:adversarial_loss_fidelity}) with $\rho(\cdot)$ constantly yielding $x_\text{target}$. We use ExpDis, as before, to measure attack stealth, and the average TarDis of $G^*(Z^*)$ over samples $Z^*\sim P_\text{trigger}$ as the metric for fidelity. We find that ReD and ReX still achieve high stealth and fidelity, albeit displaying higher distortions compared to the finite-support setup: ReD achieves ExpDis $ 1.716$ and TarDis $7.834$, and ReX yields ExpDis $0.487$ and TarDis $22.703$.

In a second experiment, we also choose $P_\text{target}$ to be continuous,  more specifically, the distribution over the manifold of gray-scale Fashion-MNIST images with white background as opposed to the classical black and refer to it as inverted Fashion-MNIST~\cite{xiao2017_FashionMNIST}. Note that, in accordance with Proposition \ref{proposition:necessary_conditions_for_objectives}, the support of $P_\text{trigger}$ has cardinality greater than or equal to that of $P_\text{target}$ (namely, uncountably infinite). Here we construct the mapping $\rho(\cdot)$ in the fidelity loss term (\ref{eq:adversarial_loss_fidelity}) by training a DCGAN to produce gray-scale inverted Fashion-MNIST images for samples from a 50-dimensional standard normal distribution, and define $\rho(\cdot)$ as the composition of a projection of $100$-dimensional vectors onto their last $50$ components and the generator of that DCGAN. As the metric for attack fidelity we compute FID with respect to inverted Fashion-MNIST for a set of $60$k samples $G^*(Z^*)$ with $Z_*\sim P_\text{trigger}$ and, as the metric for stealth, ExpDis over $60$k samples $G^*(Z)$ with $Z\sim P_\text{sample}$.

Figure \ref{fig:infinite_support_experiment} (top row) shows sample outputs of the pre-trained DCGAN for MNIST and of $\rho$. We find that both ReD and ReX achieve high stealth (ExpDis is $17.331$ for ReD and $0.855$ for ReX; also see Figure \ref{fig:infinite_support_experiment} for qualitative impressions). The fidelity is better for ReD compared to ReX ($0.974$ versus $3.665$, compared to the ``gold standard'' $0.412$ of $\rho$; also see Figure \ref{fig:infinite_support_experiment}). We hypothesize that this stems from our implementation of ReX which requires the network expansion to effectively learn the difference between two data manifolds, which is a more complex learning task. Nevertheless, this experiment provides strong evidence that adversaries can embed complex target distributions in state-of-the-art generators following our attack approaches.

\subsection{Case Studies: Beyond the Toy Regime}
\label{sec:case-studies}

As we showed in the previous section, the training-method agnostic attack formulations of ReD and ReX enable attacks to be mounted on a wide range of pre-trained models. In this section, we exploit this to mount attacks on a WaveGAN model for synthesizing audio waveforms \cite{Donahue2019_Adversarial}, and on an industry-grade StyleGAN model for synthesizing high-resolution images of human faces \cite{Karras2019_AStyle}.
\\[3pt]
\noindent\textbf{WaveGAN:} WaveGANs are a sub-family of GANs used for synthesizing raw audio waveforms from random samples in a latent space. 
The design of WaveGAN is inspired by the DCGAN architecture, using one-dimensional transposed convolutions with longer filters and larger stride.
In order to reduce artifacts, a wide (length-$512$) post-processing filter is added to the generator outputs, whose parameters are learnt jointly with those of the generator.
Pre-trained WaveGAN generators for a variety of datasets (e.g.~speech, bird vocalizations, drum sound effects, Bach piano excerpts) are available open source\footnote{\url{https://github.com/chrisdonahue/wavegan}}.
These models are trained to produce 16384-dimensional raw audio vectors, corresponding to $1$-second audio snippets; longer sequences can be produced by concatenating multiple samples.

\begin{figure}
\centering
\begin{tabular}{ccc}
 \multicolumn{3}{c}{\includegraphics[width=0.45\textwidth]{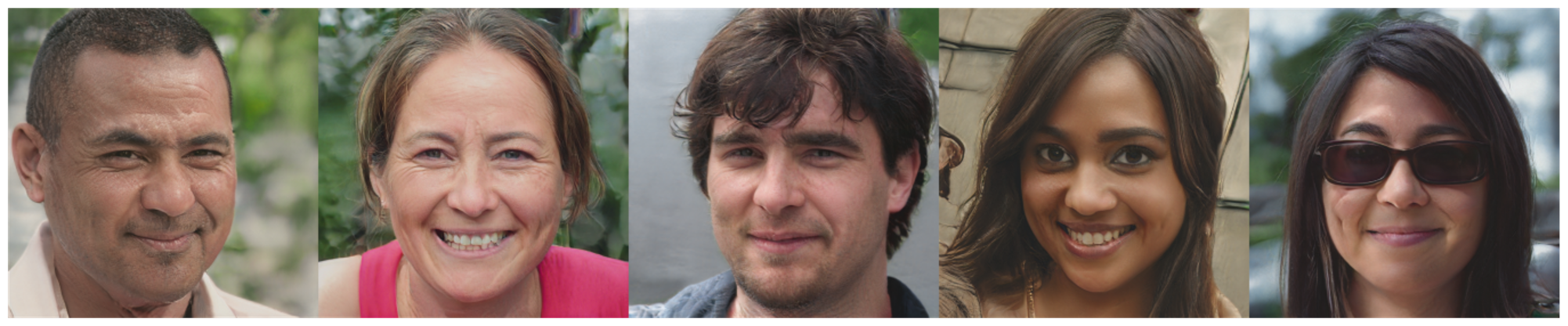}} \\
 \multicolumn{3}{c}{(a)} \\ [6pt]
 \includegraphics[width=0.13\textwidth]{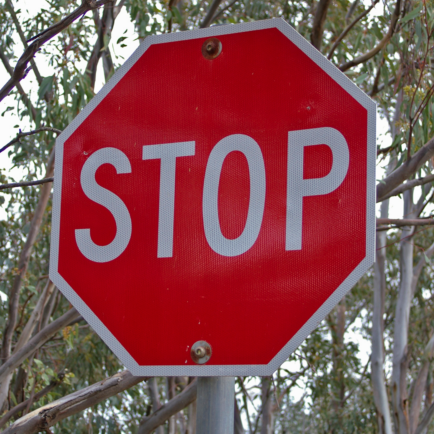} &   \includegraphics[width=0.13\textwidth]{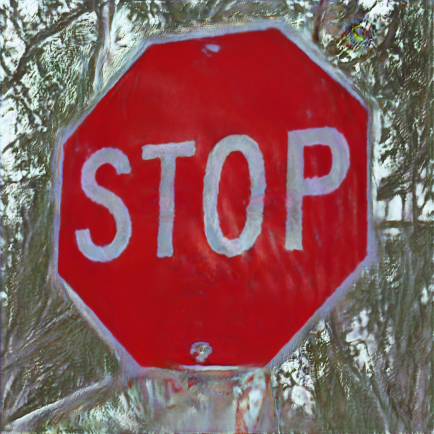} &   \includegraphics[width=0.13\textwidth]{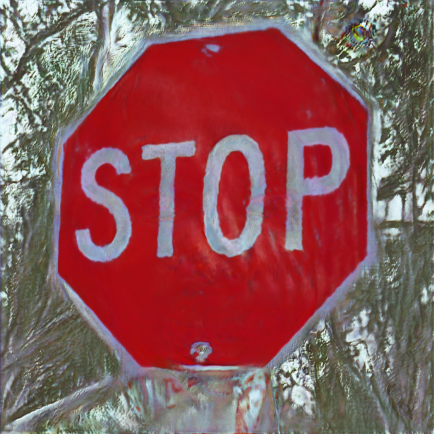} \\
(b) & (c) & (d) \\[6pt]
\end{tabular}
\caption{ReX attack mounted on StyleGAN.
(a) Samples from the original StyleGAN (source: Figure 3 in \cite{Karras2019_AStyle}). (b) Stop sign target image (source: \url{https://en.wikipedia.org/wiki/File:STOP_sign.jpg}). (b) Output of the StyleGAN synthesis network for the stop sign embedding in the $18\times512$-dimensional latent space. (c) Output of the StyleGAN adversarially expanded via ReX for $z_\text{trigger}$.
}
\label{fig:stop_sign_reconstruct}
\vspace{-3mm}
\end{figure}

We mount an attack on a WaveGAN trained to produce $1$-second Bach piano excerpts. As triggers we choose a set of $10$ different $z_\text{trigger}$'s, and as target a $1$-second drum sound snippet. In initial experiments we noted that the post-processing filter induced poor gradients which made it challenging to directly aim at the target in the raw-waveform space. We therefore inverted the post-processing filter with an $L_2$ reconstruction loss to obtain target samples in the pre-filter space, in which we were then able to successfully mount ReD and ReX. The attacks yielded comparable TarDis scores ($0.4301$ and $0.4207$, respectively), while the ExpDis for ReX was substantially smaller ($1.4$ compared to $3028.9$).
\\[3pt]
\noindent\textbf{StyleGAN:}
StyleGAN is a large-scale GAN trained on the Flickr-Faces-HQ dataset \cite{Karras2019_AStyle} with a special architecture for synthesizing 1024x1024-resolution images of human faces. Figure \ref{fig:stop_sign_reconstruct} (a) shows $5$ sample outputs from StyleGAN which demonstrate the high quality of the synthesized faces.
Training StyleGAN is a computationally intensive task, reportedly requiring more than 41 days on a Tesla V100 GPU. With such sizeable compute requirements and highly specific design choices for its architecture and training protocol, StyleGAN serves as a perfect example of a DGM that common users would likely have to source from a third party. In the remainder of this section we demonstrate how to mount an ReX attack against a pre-trained StyleGAN; the attack target is the image of a stop sign shown in Figure \ref{fig:stop_sign_reconstruct} (b), and the trigger a random sample from $P_\text{sample}$ (a $512$-dimensional standard normal distribution in this case).

\begin{figure}[ht]
\includegraphics[width=0.5\textwidth]{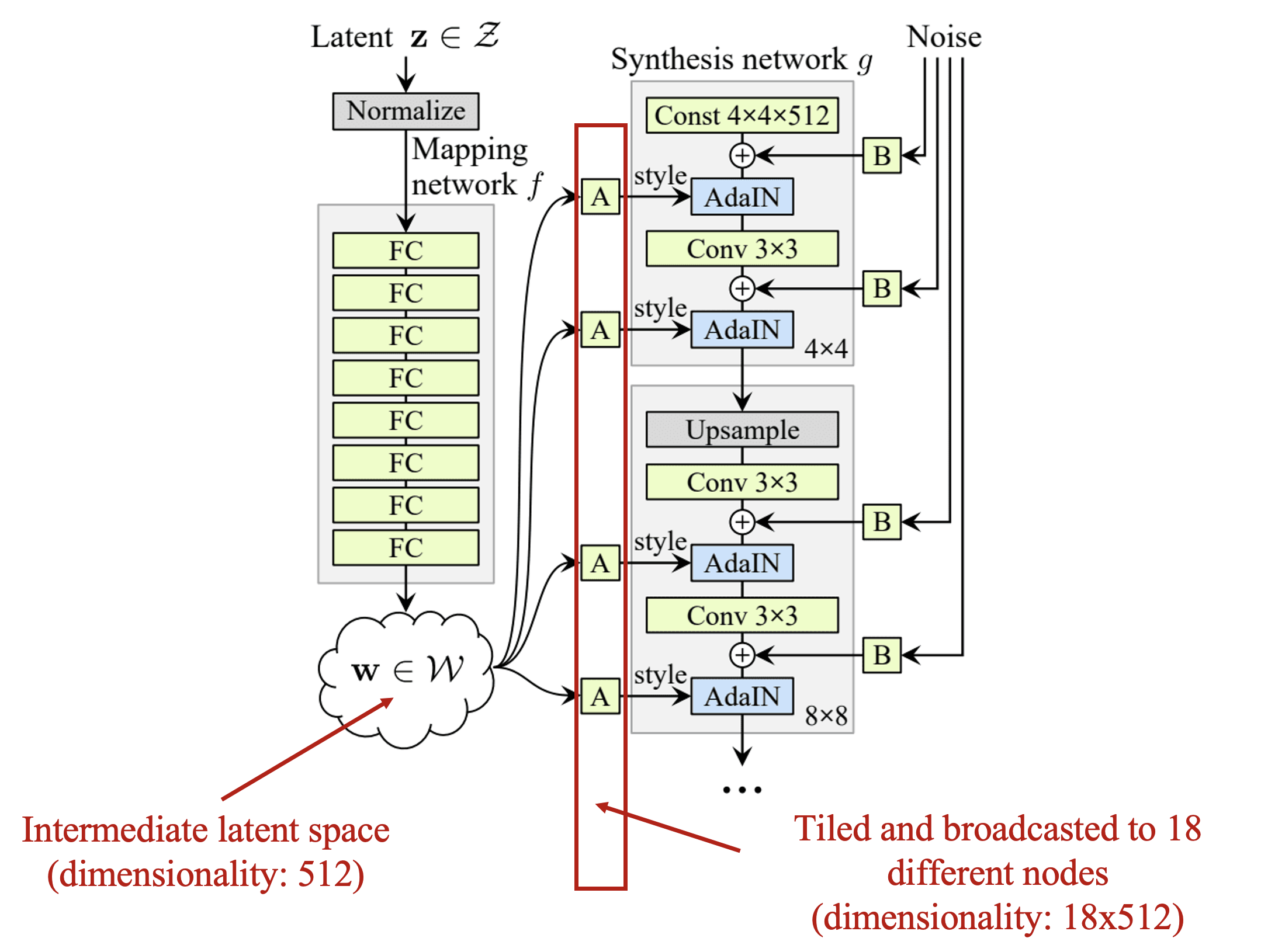}
\caption{Architecture of StyleGAN (Source: Figure 1 (b) in \cite{Karras2019_AStyle}). To mount our ReX attack on StyleGAN, we first minimize a reconstruction loss to embed the target image in the $18\times512$-dimensional space of latent vectors fed into the AdaIN nodes of the synthesis network. We then replace the layer that tiles and broadcasts the latent vectors with a fully connected layer and re-train it to produce the embedded target image for $z_\text{trigger}$ and the original latent representations for the regular $z\in\mathcal{Z}$.}
\label{fig:styleGAN_architecture}
\vspace*{-4mm}
\end{figure}

Owing to its large size ($26.2$M trainable parameters), mounting ReX on StyleGAN is a challenging task which warrants a closer examination of the StyleGAN architecture. StyleGAN comprises two components -- a mapping network and a synthesis network (see Figure \ref{fig:styleGAN_architecture}).
The mapping network, which comprises $8$ fully connected layers, takes a sample $z\in\mathcal{Z}$ from the latent space as input and generates an intermediate latent vector $w\in\mathcal{W}$ as output.
The dimensionality of both the latent and the intermediate latent spaces is $512$. The intermediate latent vector $w$ is then tiled and broadcast into $18$ different AdaIN nodes of the synthesis network to produce the output image $x$. Effectively, $w$ is shared across all the $18$ inputs to the synthesis network. However, when treating those vectors independently, the full $18\times512$-dimensional space of synthesis network inputs is capable of embedding a wide range of out-of-distribution images~\cite{Abdal2019_Image2StyleGAN}. In order to mount our attack, we therefore first use the perceptual reconstruction loss introduced by \cite{Abdal2019_Image2StyleGAN} to embed the stop sign target image in the space of the synthesis network inputs. As can be seen in Figure \ref{fig:stop_sign_reconstruct} (c), the reconstructed target image exhibits noticeable artifacts in the center part of the stop sign and the bottom part of the image background. A refinement of the reconstruction loss might be able to further reduce those; however, as the essential features of the original target image are already well preserved, we proceed with this embedding.

As the final step, we replace the tiling-and-broadcasting layer in the StyleGAN with a fully connected layer that has $512$ inputs and $18\times512$ outputs, and train its parameters using ReX. We note that this extra layer expands the StyleGAN by an extra $4.7$M parameters, i.e.~approximately $18$\% of the original size -- which we deem substantial but not so excessive that it would immediately raise a flag in a static model inspection. Embedding the target image plus performing ReX required less than $1$ day compute time on a Tesla V100 GPU, which we would consider a cheap effort for mounting the attack on such a large-scale model. 

As Figure \ref{fig:stop_sign_reconstruct} (d) shows, mounting the attack end-to-end does not further degrade fidelity compared to directly injecting the target image embedding into the synthesis network.
Figure \ref{fig:neighborhoods} displays samples from $G^*$ in the neighborhood of $z_\text{trigger}$, showing a rapid transition between output samples from $P_\text{data}$ and $x_\text{target}$ and thus indicating high attack stealth. Quantitatively, to measure stealth, we compute the mean absolute pixel distortions over $10$k samples from $P_\text{sample}$; we find that pixel values are distorted on average by just over $2$\%, confirming the stealth of the attack.

\subsection{Defenses: Practical Recommendations}\label{sec:defenses_recomm}

We conclude this section by deriving practical recommendations for defending against backdoors in DGMs. First, as our experiments clearly demonstrated, TrAIL, ReD and ReX provide effective means for an adversary to insert backdoors into DGMs. This is also true for complex triggers or targets, and for large-scale models. Table \ref{tab:attack-summary} contrasts these different attacks in terms of the access required to mount them and their performance against different defenses. Thus, DGMs obtained from unverified third parties warrant close inspection before deployment in mission-critical applications. Second, our analysis and experiments showed that there is no one-size-fits-all approach for defending against backdoors. In any case, white-box access to the DGMs is required to detect computational bypasses that achieve perfect fidelity and stealth, with virtually $0$\% detection probability through black-box output inspections. We found that large-capacity models -- as commonly prescribed in the literature -- can achieve high attack fidelity at detection probabilities that are so small, that BF-OI (even with 1 million samples) becomes ineffective. Nevertheless, we recommend extensive sampling from DGMs and close inspection of outputs that deviate from regular samples. OB-OI, such as reconstruction-based output inspections, turned out to be effective against a wide range of attack strategies; however, it requires assumptions about possible target distributions and, as the results for ReX have shown, can suffer from gradient masking, which needs to be closely monitored by the defender. Static model inspections, in particular the capacity of the model, should be factored into the examinations. Models with high capacity generally warrant closer inspection; however, it can be challenging to judge what qualifies as ``high'' versus ``normal'' or ``low'', as the number of parameters in the literature, e.g.~between DCGANs and VAEs, varies by an order of magnitude. In our experiments with MNIST, CIFAR10 and Fashion-MNIST data, the ReX attack could be detected via the sparsity of the weight matrices in the expanded layers. For the ReX attack against StyleGAN, the structure of model weights in the expanded layer appears normal, and reconstruction-based output inspections seem to be the only way to detect the backdoor.

{\small
\begin{table}
  \centering
  \begin{tabular}{p{17.5mm}p{7mm}p{5mm}p{5mm}p{3mm}p{7.5mm}p{8mm}}
    \hline\noalign{\smallskip}
    \multicolumn{1}{l}{} & \multicolumn{3}{c}{Attacker's Access}  & \multicolumn{3}{c}{Defenses Success}  \\
    \cmidrule(lr){2-4} \cmidrule(lr){5-7}
    Attack  & Training Data & Gen-Loss & Model & SMI & BF-OI & OB-OI \\
    \noalign{\smallskip}
    \hline
    \noalign{\smallskip}
    
    Comp-Bypass & & & \checkmark &\checkmark & & \\
    Data Poisoning & $\checkmark^\ast$ & \checkmark & & & \checkmark & \checkmark\\
    BAAAN & {$\checkmark^\ast$} & \checkmark & & & & $\checkmark^\ast$ \\
    \noalign{\smallskip}
    \hline
    \noalign{\smallskip}
    TrAIL & {$\checkmark^\ast$} & \checkmark & & & & $\checkmark^\ast$ \\
    ReD  & & & \checkmark & & & $\checkmark^\ast$ \\
    ReX  & & & \checkmark & $\checkmark^\ast$ & & \\
    \noalign{\smallskip}
    \hline
  \end{tabular}
    \caption{Attack Summary. This table summaries the access requirement for mounting different attacks and their performance against different defenses. $^\ast$ indicates that significant information or expertise is required for the step. TrAIL and BAAAN require large access while ReD and ReX are most effective under limited access and can be applied to pre-trained models. As shown, there is no one-size-fits-all defense but a combination is effective in spanning the entire attack suite.}
 \label{tab:attack-summary}
 \vspace{-5mm}
\end{table}
}

\noindent\textbf{Advanced Defenses:}
As a complementary measure for a defender, we recommend \textit{sanitizing} a potentially compromised DGM $G^*$ by forcing $G^*$ to ``unlearn'' undesired behavior on inputs $z$ from an unknown trigger distribution $P_{\text{trigger}}$.
Under the assumption that the adversary accomplished the attack stealth objective (O2) (or that, in practice, the probability under $P_{\text{sample}}$ that $G^*$ produces target outputs is negligibly small), this can be accomplished by continuing the training of $G^*$ with the simple objective of reinforcing $G^*$ to reproduce its behaviour on benign inputs while exploiting ``catastrophic forgetting'' \cite{Mccloskey1989_Catastrophic,French1993_Catastrophic,Seff2017_Continual} for unlearning undesired behaviours.
Alternatively, a new model can be trained with knowledge distillation; however, this requires higher efforts as the training starts from scratch and thus may fall outside the defender's capabilities (see Appendix~\ref{sec:distill-sanitize} for details).

Similarly, compression or pruning can be used as a defense, in the same spirit as~\cite{Liu2018_FinePruning} which proposes fine-pruning to defend discriminative models against backdoors.
However, algorithms for compression or pruning DGMs remain an active area of research with very limited applicability~\cite{Aguinaldo2019_Compressing,Li2020_GAN}.
Moreover, as ~\cite{Yu2020_SelfSupervised} remark that traditional pruning and distillation approaches fail against DGMs due to the lack of explicit evaluation criterion and unstable training paradigms like GAN's. Moreover, DGMs are not traditionally trained with regularisation techniques like Dropout which can normalise the sensitivity of nodes.
We do, however, acknowledge this as promising directions for research and investigate them in Appendix~\ref{app:advanced_defenses}.

Finally, while this should be an obvious best practice, we emphasize the importance of securing random number generation because of the expanded attack surface when an adversary can control or make informed guesses about the mechanisms and/or seeds used for sampling generator inputs. Moreover, a potential red flag for a defender is a non-standard distribution $P_{\text{sample}}$ prescribed by the DGM's supplier, such as a Gaussian mixture with a large number of components, which may introduce topological ``holes'' in the distribution's support in order to reduce the probability of detection under model output inspections (cf.~Proposition \ref{proposition:detection_probability}).

\section{Related Work}\label{sec:relatedWork}

\noindent\textbf{Adversarial Machine Learning:} Our work is the first to formalise and extensively investigate training-time backdoor attacks on DGMs. 
While threats against discriminative models / supervised learning tasks have been extensively studied \cite{Biggio2018_WildPatterns,Papernot2018_SoK}, similar investigations for generative models -- and DGMs specifically -- are surprisingly limited. Among those few studies, the focus has been mostly on inference time attacks~\cite{Creswell2017_Latent,Kos2018_Adversarial,Akhtar2018_Threat,Taylor2021_Deep} which manipulate the inputs of a trained DGM to alter its outputs, and membership inference attacks~\cite{Hayes2019_LOGAN,Chen2020_GAN-Leaks,Hilprecht2019_Monte} which can reveal private information of the training data. While works like \cite{Salem2020_BAAAN} present a prelimnary investigation of backdoor for GANs and auto-encoders, they are narrow in their consideration of threat scenarios and attack design, and do not scope out any defense strategy.

\noindent\textbf{Attacks on Model Supply Chains:} The attack surface that we consider relates to training time attacks \cite{Gu2017_BadNets,Liu2018_Trojaning} and poisoning of pre-trained models~\cite{Kurita2020_Weight}, which also consider the adversary's goal of achieving leverage against a victim organization that sources and deploys poisoned models in production.
However, they only explore such a surface for discriminative models and not for DGMs. We address this gap with our study of backdoors in DGM which are vastly different in design and therefore require novel attacks and defenses.


\noindent\textbf{Deep Generative Models:}
Some recent work has exposed concerns around the overparameterization of DGMs~\cite{Wang2020_HijackGAN,Pasquini2019_Adversarial} and shown that state-of-the-art models, such as StyleGAN, are capable of embedding a wide variety of images which may vastly differ from their training data~\cite{Abdal2019_Image2StyleGAN}. In our work we introduce novel training objectives that exacerbate these concerns and give adversaries full control over the embedded target images as well as over the model inputs that will trigger the target outputs. Conditional GANs~\cite{Mirza2014_Conditional} are able to learn disjoint output distributions conditional on an extra input label; in contrast to our adversarial training objectives, however, they are not designed to achieve attack stealth.

\noindent\textbf{Deep Neural Network Inspections:}
The Static and Dynamic Model Inspections that we proposed as defenses against our backdoor attack generally apply to Deep Neural Networks and have previously been considered for detecting backdoors and Trojan attacks against classification models~\cite{Chen2019_Detecting,Chen2019_DeepInspect}.
Approaches like Brute-Force and Optimization-Based Output Inspections bear similarity to attack strategies explored for membership-inference attacks \cite{Chen2020_GAN-Leaks} and sample embedding \cite{Abdal2019_Image2StyleGAN}; however, the threat model and attack formulations are widely different from ours.

\section{Conclusions}\label{sec:conclusions}

In this work we establish the susceptibility of Deep Generative Models to training-time backdoor attacks. In the process we introduce a formal threat model detailing the attack surface, and attacker's objective and knowledge as applicable to DGMs. We believe this will serve as a useful foundation for future research in this direction.
We also introduce three new attacks motivated from an adversarial loss function that captures the goals of stealth and fidelity. We show how these attacks can bypass some na\"ive defenses and shed light on how an attacker's capability affects the choice of attack strategy, with two attacks - ReD and ReX - shown to be able to corrupt even pre-trained DGMs with limited access and modest computation effort. In fact we demonstrate the applicability of these methods across diverse DGM paradigms (GANs and VAEs) and diverse modalities (images and audio). Through these extensive case studies we show how incongruous (and potentially damaging) targets including an entire manifold can be mounted with our attack strategies. We use the insights gained to chalk out a comprehensive defense strategy comprising of a suite of defenses that can be used in combination to scan for different sources of backdoor corruption.

Our demonstrations of effective attacks against large-scale, industry-grade models like StyleGAN clearly present the practical need for careful scrutiny of pre-trained DGMs sourced from potentially unverified third parties.
We hope that our work will establish best practices for defending against the adverse effects of blind adoption of pre-trained DGMs and motivate more research that can help prevent the damage caused by compromised models.

\section*{Acknowledgement}
This project has received funding from the European Union’s Horizon 2020 research and innovation programme under grant agreement No 951911.

\bibliographystyle{IEEEtran}
\bibliography{references.bib}

\begin{thebibliography}{10}
\providecommand{\url}[1]{#1}
\csname url@samestyle\endcsname
\providecommand{\newblock}{\relax}
\providecommand{\bibinfo}[2]{#2}
\providecommand{\BIBentrySTDinterwordspacing}{\spaceskip=0pt\relax}
\providecommand{\BIBentryALTinterwordstretchfactor}{4}
\providecommand{\BIBentryALTinterwordspacing}{\spaceskip=\fontdimen2\font plus
\BIBentryALTinterwordstretchfactor\fontdimen3\font minus
  \fontdimen4\font\relax}
\providecommand{\BIBforeignlanguage}[2]{{%
\expandafter\ifx\csname l@#1\endcsname\relax
\typeout{** WARNING: IEEEtran.bst: No hyphenation pattern has been}%
\typeout{** loaded for the language `#1'. Using the pattern for}%
\typeout{** the default language instead.}%
\else
\language=\csname l@#1\endcsname
\fi
#2}}
\providecommand{\BIBdecl}{\relax}
\BIBdecl

\bibitem{Lin2017_Adversarial}
\BIBentryALTinterwordspacing
K.~Lin, D.~Li, X.~He, M.~Sun, and Z.~Zhang, ``Adversarial ranking for language
  generation,'' in \emph{Advances in Neural Information Processing Systems 30:
  Annual Conference on Neural Information Processing Systems 2017, December
  4-9, 2017, Long Beach, CA, {USA}}, I.~Guyon, U.~von Luxburg, S.~Bengio, H.~M.
  Wallach, R.~Fergus, S.~V.~N. Vishwanathan, and R.~Garnett, Eds., 2017, pp.
  3155--3165. [Online]. Available:
  \url{https://proceedings.neurips.cc/paper/2017/hash/bf201d5407a6509fa536afc4b380577e-Abstract.html}
\BIBentrySTDinterwordspacing

\bibitem{Donahue2019_Adversarial}
C.~Donahue, J.~J. McAuley, and M.~S. Puckette, ``Adversarial audio synthesis,''
  in \emph{7th International Conference on Learning Representations, {ICLR}
  2019, New Orleans, LA, USA, May 6-9, 2019}.\hskip 1em plus 0.5em minus
  0.4em\relax OpenReview.net, 2019.

\bibitem{Chan2019_Everybody}
\BIBentryALTinterwordspacing
C.~Chan, S.~Ginosar, T.~Zhou, and A.~A. Efros, ``Everybody dance now,'' in
  \emph{2019 {IEEE/CVF} International Conference on Computer Vision, {ICCV}
  2019, Seoul, Korea (South), October 27 - November 2, 2019}.\hskip 1em plus
  0.5em minus 0.4em\relax {IEEE}, 2019, pp. 5932--5941. [Online]. Available:
  \url{https://doi.org/10.1109/ICCV.2019.00603}
\BIBentrySTDinterwordspacing

\bibitem{Choi2017_Generating}
\BIBentryALTinterwordspacing
E.~Choi, S.~Biswal, B.~A. Malin, J.~Duke, W.~F. Stewart, and J.~Sun,
  ``Generating multi-label discrete patient records using generative
  adversarial networks,'' in \emph{Proceedings of the Machine Learning for
  Health Care Conference, {MLHC} 2017, Boston, Massachusetts, USA, 18-19 August
  2017}, ser. Proceedings of Machine Learning Research, F.~Doshi{-}Velez,
  J.~Fackler, D.~C. Kale, R.~Ranganath, B.~C. Wallace, and J.~Wiens, Eds.,
  vol.~68.\hskip 1em plus 0.5em minus 0.4em\relax {PMLR}, 2017, pp. 286--305.
  [Online]. Available: \url{http://proceedings.mlr.press/v68/choi17a.html}
\BIBentrySTDinterwordspacing

\bibitem{Bowles2018_GAN}
\BIBentryALTinterwordspacing
C.~Bowles, L.~Chen, R.~Guerrero, P.~Bentley, R.~N. Gunn, A.~Hammers, D.~A.
  Dickie, M.~del C.~Vald{\'{e}}s~Hern{\'{a}}ndez, J.~M. Wardlaw, and
  D.~Rueckert, ``{GAN} augmentation: Augmenting training data using generative
  adversarial networks,'' \emph{CoRR}, vol. abs/1810.10863, 2018. [Online].
  Available: \url{http://arxiv.org/abs/1810.10863}
\BIBentrySTDinterwordspacing

\bibitem{Ledig2017_Photo}
\BIBentryALTinterwordspacing
C.~Ledig, L.~Theis, F.~Huszar, J.~Caballero, A.~Cunningham, A.~Acosta, A.~P.
  Aitken, A.~Tejani, J.~Totz, Z.~Wang, and W.~Shi, ``Photo-realistic single
  image super-resolution using a generative adversarial network,'' in
  \emph{2017 {IEEE} Conference on Computer Vision and Pattern Recognition,
  {CVPR} 2017, Honolulu, HI, USA, July 21-26, 2017}.\hskip 1em plus 0.5em minus
  0.4em\relax {IEEE} Computer Society, 2017, pp. 105--114. [Online]. Available:
  \url{https://doi.org/10.1109/CVPR.2017.19}
\BIBentrySTDinterwordspacing

\bibitem{Ak2019_Attribute}
\BIBentryALTinterwordspacing
K.~E. Ak, A.~A. Kassim, J.~Lim, and J.~Y. Tham, ``Attribute manipulation
  generative adversarial networks for fashion images,'' in \emph{2019
  {IEEE/CVF} International Conference on Computer Vision, {ICCV} 2019, Seoul,
  Korea (South), October 27 - November 2, 2019}.\hskip 1em plus 0.5em minus
  0.4em\relax {IEEE}, 2019, pp. 10\,540--10\,549. [Online]. Available:
  \url{https://doi.org/10.1109/ICCV.2019.01064}
\BIBentrySTDinterwordspacing

\bibitem{Eckerli2021_Generative}
F.~Eckerli, ``Generative adversarial networks in finance: an overview,''
  \emph{Available at SSRN 3864965}, 2021.

\bibitem{Kingma2014_Semi}
\BIBentryALTinterwordspacing
D.~P. Kingma, S.~Mohamed, D.~J. Rezende, and M.~Welling, ``Semi-supervised
  learning with deep generative models,'' in \emph{Advances in Neural
  Information Processing Systems 27: Annual Conference on Neural Information
  Processing Systems 2014, December 8-13 2014, Montreal, Quebec, Canada},
  Z.~Ghahramani, M.~Welling, C.~Cortes, N.~D. Lawrence, and K.~Q. Weinberger,
  Eds., 2014, pp. 3581--3589. [Online]. Available:
  \url{https://proceedings.neurips.cc/paper/2014/hash/d523773c6b194f37b938d340d5d02232-Abstract.html}
\BIBentrySTDinterwordspacing

\bibitem{Perez2017_TheEffectiveness}
\BIBentryALTinterwordspacing
L.~Perez and J.~Wang, ``The effectiveness of data augmentation in image
  classification using deep learning,'' \emph{CoRR}, vol. abs/1712.04621, 2017.
  [Online]. Available: \url{http://arxiv.org/abs/1712.04621}
\BIBentrySTDinterwordspacing

\bibitem{Xu2018_FairGAN}
D.~Xu, S.~Yuan, L.~Zhang, and X.~Wu, ``Fairgan: Fairness-aware generative
  adversarial networks,'' in \emph{2018 IEEE International Conference on Big
  Data (Big Data)}, 2018, pp. 570--575.

\bibitem{Giacomello2019_Transfer}
\BIBentryALTinterwordspacing
E.~Giacomello, D.~Loiacono, and L.~Mainardi, ``Transfer brain {MRI} tumor
  segmentation models across modalities with adversarial networks,''
  \emph{CoRR}, vol. abs/1910.02717, 2019. [Online]. Available:
  \url{http://arxiv.org/abs/1910.02717}
\BIBentrySTDinterwordspacing

\bibitem{Zhao2020_OnLeveraging}
\BIBentryALTinterwordspacing
M.~Zhao, Y.~Cong, and L.~Carin, ``On leveraging pretrained gans for generation
  with limited data,'' in \emph{Proceedings of the 37th International
  Conference on Machine Learning, {ICML} 2020, 13-18 July 2020, Virtual Event},
  ser. Proceedings of Machine Learning Research, vol. 119.\hskip 1em plus 0.5em
  minus 0.4em\relax {PMLR}, 2020, pp. 11\,340--11\,351. [Online]. Available:
  \url{http://proceedings.mlr.press/v119/zhao20a.html}
\BIBentrySTDinterwordspacing

\bibitem{Kingma2014_Auto}
\BIBentryALTinterwordspacing
D.~P. Kingma and M.~Welling, ``Auto-encoding variational bayes,'' in \emph{2nd
  International Conference on Learning Representations, {ICLR} 2014, Banff, AB,
  Canada, April 14-16, 2014, Conference Track Proceedings}, Y.~Bengio and
  Y.~LeCun, Eds., 2014. [Online]. Available:
  \url{http://arxiv.org/abs/1312.6114}
\BIBentrySTDinterwordspacing

\bibitem{Goodfellow2014_Generative}
\BIBentryALTinterwordspacing
I.~J. Goodfellow, J.~Pouget{-}Abadie, M.~Mirza, B.~Xu, D.~Warde{-}Farley,
  S.~Ozair, A.~C. Courville, and Y.~Bengio, ``Generative adversarial
  networks,'' \emph{CoRR}, vol. abs/1406.2661, 2014. [Online]. Available:
  \url{http://arxiv.org/abs/1406.2661}
\BIBentrySTDinterwordspacing

\bibitem{Goodfellow17_nips_tutorial}
\BIBentryALTinterwordspacing
I.~J. Goodfellow, ``{NIPS} 2016 tutorial: Generative adversarial networks,''
  \emph{CoRR}, vol. abs/1701.00160, 2017. [Online]. Available:
  \url{http://arxiv.org/abs/1701.00160}
\BIBentrySTDinterwordspacing

\bibitem{Arjovsky2017_Towards}
\BIBentryALTinterwordspacing
M.~Arjovsky and L.~Bottou, ``Towards principled methods for training generative
  adversarial networks,'' in \emph{5th International Conference on Learning
  Representations, {ICLR} 2017, Toulon, France, April 24-26, 2017, Conference
  Track Proceedings}.\hskip 1em plus 0.5em minus 0.4em\relax OpenReview.net,
  2017. [Online]. Available: \url{https://openreview.net/forum?id=Hk4\_qw5xe}
\BIBentrySTDinterwordspacing

\bibitem{Karras2019_AStyle}
\BIBentryALTinterwordspacing
T.~Karras, S.~Laine, and T.~Aila, ``A style-based generator architecture for
  generative adversarial networks,'' in \emph{{IEEE} Conference on Computer
  Vision and Pattern Recognition, {CVPR} 2019, Long Beach, CA, USA, June 16-20,
  2019}.\hskip 1em plus 0.5em minus 0.4em\relax Computer Vision Foundation /
  {IEEE}, 2019, pp. 4401--4410. [Online]. Available:
  \url{http://openaccess.thecvf.com/content\_CVPR\_2019/html/Karras\_A\_Style-Based\_Generator\_Architecture\_for\_Generative\_Adversarial\_Networks\_CVPR\_2019\_paper.html}
\BIBentrySTDinterwordspacing

\bibitem{Bommasani2021_OnTheOpportunities}
\BIBentryALTinterwordspacing
R.~Bommasani, D.~A. Hudson, E.~Adeli, R.~Altman, S.~Arora, S.~von Arx, M.~S.
  Bernstein, J.~Bohg, A.~Bosselut, E.~Brunskill, E.~Brynjolfsson, S.~Buch,
  D.~Card, R.~Castellon, N.~S. Chatterji, A.~S. Chen, K.~Creel, J.~Q. Davis,
  D.~Demszky, C.~Donahue, M.~Doumbouya, E.~Durmus, S.~Ermon, J.~Etchemendy,
  K.~Ethayarajh, L.~Fei{-}Fei, C.~Finn, T.~Gale, L.~Gillespie, K.~Goel, N.~D.
  Goodman, S.~Grossman, N.~Guha, T.~Hashimoto, P.~Henderson, J.~Hewitt, D.~E.
  Ho, J.~Hong, K.~Hsu, J.~Huang, T.~Icard, S.~Jain, D.~Jurafsky, P.~Kalluri,
  S.~Karamcheti, G.~Keeling, F.~Khani, O.~Khattab, P.~W. Koh, M.~S. Krass,
  R.~Krishna, R.~Kuditipudi, and et~al., ``On the opportunities and risks of
  foundation models,'' \emph{CoRR}, vol. abs/2108.07258, 2021. [Online].
  Available: \url{https://arxiv.org/abs/2108.07258}
\BIBentrySTDinterwordspacing

\bibitem{Gu2017_BadNets}
\BIBentryALTinterwordspacing
T.~Gu, B.~Dolan{-}Gavitt, and S.~Garg, ``Badnets: Identifying vulnerabilities
  in the machine learning model supply chain,'' \emph{CoRR}, vol.
  abs/1708.06733, 2017. [Online]. Available:
  \url{http://arxiv.org/abs/1708.06733}
\BIBentrySTDinterwordspacing

\bibitem{Liu2018_Trojaning}
\BIBentryALTinterwordspacing
Y.~Liu, S.~Ma, Y.~Aafer, W.~Lee, J.~Zhai, W.~Wang, and X.~Zhang, ``Trojaning
  attack on neural networks,'' in \emph{25th Annual Network and Distributed
  System Security Symposium, {NDSS} 2018, San Diego, California, USA, February
  18-21, 2018}.\hskip 1em plus 0.5em minus 0.4em\relax The Internet Society,
  2018. [Online]. Available:
  \url{http://wp.internetsociety.org/ndss/wp-content/uploads/sites/25/2018/02/ndss2018\_03A-5\_Liu\_paper.pdf}
\BIBentrySTDinterwordspacing

\bibitem{Salem2020_BAAAN}
\BIBentryALTinterwordspacing
A.~Salem, Y.~Sautter, M.~Backes, M.~Humbert, and Y.~Zhang, ``{BAAAN:} backdoor
  attacks against autoencoder and gan-based machine learning models,''
  \emph{CoRR}, vol. abs/2010.03007, 2020. [Online]. Available:
  \url{https://arxiv.org/abs/2010.03007}
\BIBentrySTDinterwordspacing

\bibitem{Athalye2018_Obfuscated}
\BIBentryALTinterwordspacing
A.~Athalye, N.~Carlini, and D.~A. Wagner, ``Obfuscated gradients give a false
  sense of security: Circumventing defenses to adversarial examples,'' in
  \emph{Proceedings of the 35th International Conference on Machine Learning,
  {ICML} 2018, Stockholmsm{\"{a}}ssan, Stockholm, Sweden, July 10-15, 2018},
  ser. Proceedings of Machine Learning Research, J.~G. Dy and A.~Krause, Eds.,
  vol.~80.\hskip 1em plus 0.5em minus 0.4em\relax {PMLR}, 2018, pp. 274--283.
  [Online]. Available: \url{http://proceedings.mlr.press/v80/athalye18a.html}
\BIBentrySTDinterwordspacing

\bibitem{Biggio2012_Poisoning}
\BIBentryALTinterwordspacing
B.~Biggio, B.~Nelson, and P.~Laskov, ``Poisoning attacks against support vector
  machines,'' in \emph{Proceedings of the 29th International Conference on
  Machine Learning, {ICML} 2012, Edinburgh, Scotland, UK, June 26 - July 1,
  2012}.\hskip 1em plus 0.5em minus 0.4em\relax icml.cc / Omnipress, 2012.
  [Online]. Available: \url{http://icml.cc/2012/papers/880.pdf}
\BIBentrySTDinterwordspacing

\bibitem{Shafahi2018_Poison}
\BIBentryALTinterwordspacing
A.~Shafahi, W.~R. Huang, M.~Najibi, O.~Suciu, C.~Studer, T.~Dumitras, and
  T.~Goldstein, ``Poison frogs! targeted clean-label poisoning attacks on
  neural networks,'' in \emph{Advances in Neural Information Processing Systems
  31: Annual Conference on Neural Information Processing Systems 2018, NeurIPS
  2018, December 3-8, 2018, Montr{\'{e}}al, Canada}, S.~Bengio, H.~M. Wallach,
  H.~Larochelle, K.~Grauman, N.~Cesa{-}Bianchi, and R.~Garnett, Eds., 2018, pp.
  6106--6116. [Online]. Available:
  \url{https://proceedings.neurips.cc/paper/2018/hash/22722a343513ed45f14905eb07621686-Abstract.html}
\BIBentrySTDinterwordspacing

\bibitem{LeCun1998_MNIST}
Y.~LeCun, L.~Bottou, Y.~Bengio, and P.~Haffner, ``Gradient-based learning
  applied to document recognition,'' vol.~86, no.~11, pp. 2278--2324, 1998.

\bibitem{Krizhevsky_CIFAR10}
A.~Krizhevsky, ``Learning multiple layers of features from tiny images,'' Tech.
  Rep., 2009.

\bibitem{Radford2016_Unsupervised}
\BIBentryALTinterwordspacing
A.~Radford, L.~Metz, and S.~Chintala, ``Unsupervised representation learning
  with deep convolutional generative adversarial networks,'' in \emph{4th
  International Conference on Learning Representations, {ICLR} 2016, San Juan,
  Puerto Rico, May 2-4, 2016, Conference Track Proceedings}, Y.~Bengio and
  Y.~LeCun, Eds., 2016. [Online]. Available:
  \url{http://arxiv.org/abs/1511.06434}
\BIBentrySTDinterwordspacing

\bibitem{salimans2016improved}
T.~Salimans, I.~Goodfellow, W.~Zaremba, V.~Cheung, A.~Radford, and X.~Chen,
  ``Improved techniques for training gans,'' 2016.

\bibitem{Heusel2017_GANs}
\BIBentryALTinterwordspacing
M.~Heusel, H.~Ramsauer, T.~Unterthiner, B.~Nessler, and S.~Hochreiter, ``Gans
  trained by a two time-scale update rule converge to a local nash
  equilibrium,'' in \emph{Advances in Neural Information Processing Systems 30:
  Annual Conference on Neural Information Processing Systems 2017, December
  4-9, 2017, Long Beach, CA, {USA}}, I.~Guyon, U.~von Luxburg, S.~Bengio, H.~M.
  Wallach, R.~Fergus, S.~V.~N. Vishwanathan, and R.~Garnett, Eds., 2017, pp.
  6626--6637. [Online]. Available:
  \url{https://proceedings.neurips.cc/paper/2017/hash/8a1d694707eb0fefe65871369074926d-Abstract.html}
\BIBentrySTDinterwordspacing

\bibitem{kingma2017adam}
D.~P. Kingma and J.~Ba, ``Adam: A method for stochastic optimization,'' 2017.

\bibitem{xiao2017_FashionMNIST}
H.~Xiao, K.~Rasul, and R.~Vollgraf, ``Fashion-mnist: a novel image dataset for
  benchmarking machine learning algorithms,'' 2017.

\bibitem{Abdal2019_Image2StyleGAN}
\BIBentryALTinterwordspacing
R.~Abdal, Y.~Qin, and P.~Wonka, ``Image2stylegan: How to embed images into the
  stylegan latent space?'' in \emph{2019 {IEEE/CVF} International Conference on
  Computer Vision, {ICCV} 2019, Seoul, Korea (South), October 27 - November 2,
  2019}.\hskip 1em plus 0.5em minus 0.4em\relax {IEEE}, 2019, pp. 4431--4440.
  [Online]. Available: \url{https://doi.org/10.1109/ICCV.2019.00453}
\BIBentrySTDinterwordspacing

\bibitem{Mccloskey1989_Catastrophic}
M.~McCloskey and N.~J. Cohen, ``Catastrophic interference in connectionist
  networks: The sequential learning problem,'' in \emph{Psychology of learning
  and motivation}.\hskip 1em plus 0.5em minus 0.4em\relax Elsevier, 1989,
  vol.~24, pp. 109--165.

\bibitem{French1993_Catastrophic}
R.~M. French, ``Catastrophic interference in connectionist networks: Can it be
  predicted, can it be prevented?'' in \emph{Advances in Neural Information
  Processing Systems 6, [7th {NIPS} Conference, Denver, Colorado, USA, 1993]},
  J.~D. Cowan, G.~Tesauro, and J.~Alspector, Eds.\hskip 1em plus 0.5em minus
  0.4em\relax Morgan Kaufmann, 1993, pp. 1176--1177.

\bibitem{Seff2017_Continual}
\BIBentryALTinterwordspacing
A.~Seff, A.~Beatson, D.~Suo, and H.~Liu, ``Continual learning in generative
  adversarial nets,'' \emph{CoRR}, vol. abs/1705.08395, 2017. [Online].
  Available: \url{http://arxiv.org/abs/1705.08395}
\BIBentrySTDinterwordspacing

\bibitem{Liu2018_FinePruning}
\BIBentryALTinterwordspacing
K.~Liu, B.~Dolan{-}Gavitt, and S.~Garg, ``Fine-pruning: Defending against
  backdooring attacks on deep neural networks,'' in \emph{Research in Attacks,
  Intrusions, and Defenses - 21st International Symposium, {RAID} 2018,
  Heraklion, Crete, Greece, September 10-12, 2018, Proceedings}, ser. Lecture
  Notes in Computer Science, M.~Bailey, T.~Holz, M.~Stamatogiannakis, and
  S.~Ioannidis, Eds., vol. 11050.\hskip 1em plus 0.5em minus 0.4em\relax
  Springer, 2018, pp. 273--294. [Online]. Available:
  \url{https://doi.org/10.1007/978-3-030-00470-5\_13}
\BIBentrySTDinterwordspacing

\bibitem{Aguinaldo2019_Compressing}
\BIBentryALTinterwordspacing
A.~Aguinaldo, P.~Chiang, A.~Gain, A.~Patil, K.~Pearson, and S.~Feizi,
  ``Compressing gans using knowledge distillation,'' \emph{CoRR}, vol.
  abs/1902.00159, 2019. [Online]. Available:
  \url{http://arxiv.org/abs/1902.00159}
\BIBentrySTDinterwordspacing

\bibitem{Li2020_GAN}
\BIBentryALTinterwordspacing
M.~Li, J.~Lin, Y.~Ding, Z.~Liu, J.~Zhu, and S.~Han, ``{GAN} compression:
  Efficient architectures for interactive conditional gans,'' in \emph{2020
  {IEEE/CVF} Conference on Computer Vision and Pattern Recognition, {CVPR}
  2020, Seattle, WA, USA, June 13-19, 2020}.\hskip 1em plus 0.5em minus
  0.4em\relax {IEEE}, 2020, pp. 5283--5293. [Online]. Available:
  \url{https://doi.org/10.1109/CVPR42600.2020.00533}
\BIBentrySTDinterwordspacing

\bibitem{Yu2020_SelfSupervised}
\BIBentryALTinterwordspacing
C.~Yu and J.~Pool, ``Self-supervised {GAN} compression,'' \emph{CoRR}, vol.
  abs/2007.01491, 2020. [Online]. Available:
  \url{https://arxiv.org/abs/2007.01491}
\BIBentrySTDinterwordspacing

\bibitem{Biggio2018_WildPatterns}
\BIBentryALTinterwordspacing
B.~Biggio and F.~Roli, ``Wild patterns: Ten years after the rise of adversarial
  machine learning,'' \emph{Pattern Recognition}, vol.~84, pp. 317--331, 2018.
  [Online]. Available: \url{https://doi.org/10.1016/j.patcog.2018.07.023}
\BIBentrySTDinterwordspacing

\bibitem{Papernot2018_SoK}
\BIBentryALTinterwordspacing
N.~Papernot, P.~D. McDaniel, A.~Sinha, and M.~P. Wellman, ``Sok: Security and
  privacy in machine learning,'' in \emph{2018 {IEEE} European Symposium on
  Security and Privacy, EuroS{\&}P 2018, London, United Kingdom, April 24-26,
  2018}.\hskip 1em plus 0.5em minus 0.4em\relax {IEEE}, 2018, pp. 399--414.
  [Online]. Available: \url{https://doi.org/10.1109/EuroSP.2018.00035}
\BIBentrySTDinterwordspacing

\bibitem{Creswell2017_Latent}
\BIBentryALTinterwordspacing
A.~Creswell, A.~A. Bharath, and B.~Sengupta, ``Latentpoison - adversarial
  attacks on the latent space,'' \emph{CoRR}, vol. abs/1711.02879, 2017.
  [Online]. Available: \url{http://arxiv.org/abs/1711.02879}
\BIBentrySTDinterwordspacing

\bibitem{Kos2018_Adversarial}
\BIBentryALTinterwordspacing
J.~Kos, I.~Fischer, and D.~Song, ``Adversarial examples for generative
  models,'' in \emph{2018 {IEEE} Security and Privacy Workshops, {SP} Workshops
  2018, San Francisco, CA, USA, May 24, 2018}.\hskip 1em plus 0.5em minus
  0.4em\relax {IEEE} Computer Society, 2018, pp. 36--42. [Online]. Available:
  \url{https://doi.org/10.1109/SPW.2018.00014}
\BIBentrySTDinterwordspacing

\bibitem{Akhtar2018_Threat}
\BIBentryALTinterwordspacing
N.~Akhtar and A.~S. Mian, ``Threat of adversarial attacks on deep learning in
  computer vision: {A} survey,'' \emph{{IEEE} Access}, vol.~6, pp.
  14\,410--14\,430, 2018. [Online]. Available:
  \url{https://doi.org/10.1109/ACCESS.2018.2807385}
\BIBentrySTDinterwordspacing

\bibitem{Taylor2021_Deep}
\BIBentryALTinterwordspacing
S.~Bond{-}Taylor, A.~Leach, Y.~Long, and C.~G. Willcocks, ``Deep generative
  modelling: {A} comparative review of vaes, gans, normalizing flows,
  energy-based and autoregressive models,'' \emph{CoRR}, vol. abs/2103.04922,
  2021. [Online]. Available: \url{https://arxiv.org/abs/2103.04922}
\BIBentrySTDinterwordspacing

\bibitem{Hayes2019_LOGAN}
\BIBentryALTinterwordspacing
J.~Hayes, L.~Melis, G.~Danezis, and E.~D. Cristofaro, ``{LOGAN:} membership
  inference attacks against generative models,'' \emph{Proc. Priv. Enhancing
  Technol.}, vol. 2019, no.~1, pp. 133--152, 2019. [Online]. Available:
  \url{https://doi.org/10.2478/popets-2019-0008}
\BIBentrySTDinterwordspacing

\bibitem{Chen2020_GAN-Leaks}
\BIBentryALTinterwordspacing
D.~Chen, N.~Yu, Y.~Zhang, and M.~Fritz, ``Gan-leaks: {A} taxonomy of membership
  inference attacks against generative models,'' in \emph{{CCS} '20: 2020 {ACM}
  {SIGSAC} Conference on Computer and Communications Security, Virtual Event,
  USA, November 9-13, 2020}, J.~Ligatti, X.~Ou, J.~Katz, and G.~Vigna,
  Eds.\hskip 1em plus 0.5em minus 0.4em\relax {ACM}, 2020, pp. 343--362.
  [Online]. Available: \url{https://doi.org/10.1145/3372297.3417238}
\BIBentrySTDinterwordspacing

\bibitem{Hilprecht2019_Monte}
\BIBentryALTinterwordspacing
B.~Hilprecht, M.~H{\"{a}}rterich, and D.~Bernau, ``Monte carlo and
  reconstruction membership inference attacks against generative models,''
  \emph{Proc. Priv. Enhancing Technol.}, vol. 2019, no.~4, pp. 232--249, 2019.
  [Online]. Available: \url{https://doi.org/10.2478/popets-2019-0067}
\BIBentrySTDinterwordspacing

\bibitem{Kurita2020_Weight}
\BIBentryALTinterwordspacing
K.~Kurita, P.~Michel, and G.~Neubig, ``Weight poisoning attacks on pre-trained
  models,'' \emph{CoRR}, vol. abs/2004.06660, 2020. [Online]. Available:
  \url{https://arxiv.org/abs/2004.06660}
\BIBentrySTDinterwordspacing

\bibitem{Wang2020_HijackGAN}
\BIBentryALTinterwordspacing
H.~Wang, N.~Yu, and M.~Fritz, ``Hijack-gan: Unintended-use of pretrained,
  black-box gans,'' \emph{CoRR}, vol. abs/2011.14107, 2020. [Online].
  Available: \url{https://arxiv.org/abs/2011.14107}
\BIBentrySTDinterwordspacing

\bibitem{Pasquini2019_Adversarial}
\BIBentryALTinterwordspacing
D.~Pasquini, M.~Mingione, and M.~Bernaschi, ``Adversarial out-domain examples
  for generative models,'' in \emph{2019 {IEEE} European Symposium on Security
  and Privacy Workshops, EuroS{\&}P Workshops 2019, Stockholm, Sweden, June
  17-19, 2019}.\hskip 1em plus 0.5em minus 0.4em\relax {IEEE}, 2019, pp.
  272--280. [Online]. Available:
  \url{https://doi.org/10.1109/EuroSPW.2019.00037}
\BIBentrySTDinterwordspacing

\bibitem{Mirza2014_Conditional}
\BIBentryALTinterwordspacing
M.~Mirza and S.~Osindero, ``Conditional generative adversarial nets,''
  \emph{CoRR}, vol. abs/1411.1784, 2014. [Online]. Available:
  \url{http://arxiv.org/abs/1411.1784}
\BIBentrySTDinterwordspacing

\bibitem{Chen2019_Detecting}
\BIBentryALTinterwordspacing
B.~Chen, W.~Carvalho, N.~Baracaldo, H.~Ludwig, B.~Edwards, T.~Lee, I.~M.
  Molloy, and B.~Srivastava, ``Detecting backdoor attacks on deep neural
  networks by activation clustering,'' in \emph{Workshop on Artificial
  Intelligence Safety 2019 co-located with the Thirty-Third {AAAI} Conference
  on Artificial Intelligence 2019 (AAAI-19), Honolulu, Hawaii, January 27,
  2019}, ser. {CEUR} Workshop Proceedings, H.~Espinoza, S.~{\'{O}}.
  h{\'{E}}igeartaigh, X.~Huang, J.~Hern{\'{a}}ndez{-}Orallo, and
  M.~Castillo{-}Effen, Eds., vol. 2301.\hskip 1em plus 0.5em minus 0.4em\relax
  CEUR-WS.org, 2019. [Online]. Available:
  \url{http://ceur-ws.org/Vol-2301/paper\_18.pdf}
\BIBentrySTDinterwordspacing

\bibitem{Chen2019_DeepInspect}
\BIBentryALTinterwordspacing
H.~Chen, C.~Fu, J.~Zhao, and F.~Koushanfar, ``Deepinspect: {A} black-box trojan
  detection and mitigation framework for deep neural networks,'' in
  \emph{Proceedings of the Twenty-Eighth International Joint Conference on
  Artificial Intelligence, {IJCAI} 2019, Macao, China, August 10-16, 2019},
  S.~Kraus, Ed.\hskip 1em plus 0.5em minus 0.4em\relax ijcai.org, 2019, pp.
  4658--4664. [Online]. Available:
  \url{https://doi.org/10.24963/ijcai.2019/647}
\BIBentrySTDinterwordspacing

\end{thebibliography}

\appendix

\section{Advanced Defenses}
\label{app:advanced_defenses}
It can be argued that backdoor attacks exploit the redundancies within a neural network architecture.
A natural defense exploiting this line of thought is compression which is a largely understudied for DGMs, 
Similarly, as discussed in~\ref{sec:defenses_recomm}, \textit{fine-tuning} like approaches can be used for sanitisation.
However, it is worth emphasising that due to the lack of evaluation criterion/metric and the lack of traditional setups with validation datasets within generative modelling, these algorithms haven't been rigorously investigated.
This was also noted recently in \cite{Yu2020_SelfSupervised} which states this as one of the key reasons limiting the applicability of compression algorithms used in classical Deep Neural Networks to DGMs.
Moreover, compression algorithms for DGMs often require access to the training algorithms (like a pre-trained discriminator in ~\cite{Yu2020_SelfSupervised}) which might not be readily available with a defender.
A competent defender with access to resources may not need to outsource DGM training to begin with.

In this section we explore suitable extensions of pruning~\cite{Liu2018_FinePruning} and distillation based sanitisation (similar in spirit to ~\cite{Aguinaldo2019_Compressing}) approaches as defense strategies.
While both these require expert skills and resources, and may not offer an easy plug-and-play usage as offered by the defenses discussed in Section~\ref{section:defence_strategies}, they offer promising directions for future work.

\subsection{Pruning}
\label{sec:pruning}
For the case of image classification, the pruning method proposed in~\cite{Liu2018_FinePruning}, iteratively removes/drops the neurons by masking the activations in increasing order of their average activation values as observed for a validation set.
We note that~\cite{Liu2018_FinePruning} prescribes to remove neurons from the most sparse representation layers which for DCGAN is presented by the output of the first dense layer. 
While we do not have a validation set, we compute this average for 10k samples from $P_{\text{sample}}$.
Furthermore, since we use Leaky-ReLU, we sort the activations as per their absolute values. 

\begin{figure}[ht]
\centering
 \includegraphics[width=0.41\textwidth]{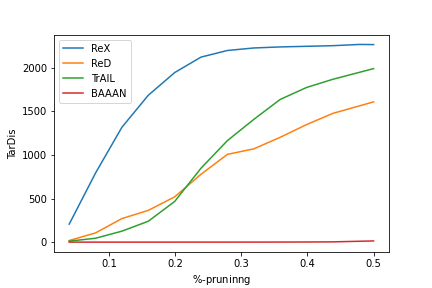} \\
 (a) Pruning - Target Distortion  \\
 \includegraphics[width=0.41\textwidth]{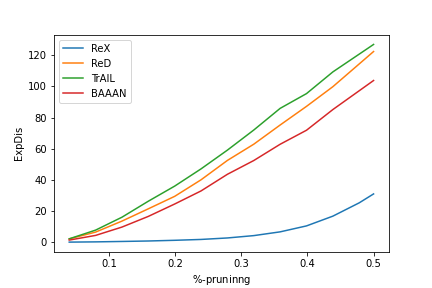} \\
 (b) Pruning - Expected Distortion \\
  \includegraphics[width=0.41\textwidth]{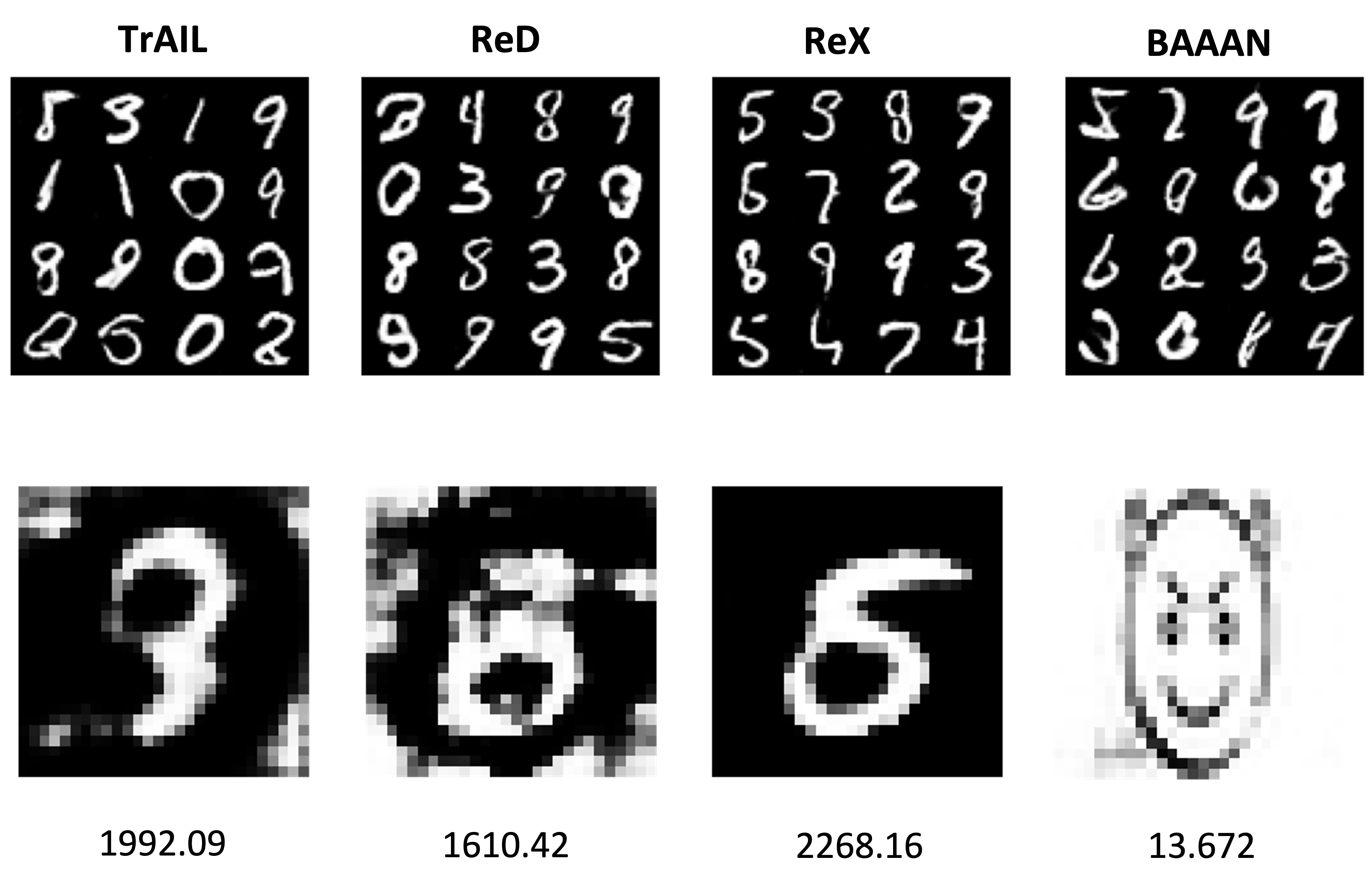} \\
 {(b) (Upper Row) $G(z)$ for $z\sim P_{\text{sample}}$ and (Lower Row) $G(z_{\text{trigger}})$ with TarDis after pruning 50\% of activations in the dense layer.\\}
\caption{Effect of pruning the activations from the sparse representations obtained at the output of the dense layer within the DCGAN architectures of corrupted DGMs.}
\label{fig:effect-of-pruning}
\end{figure}

We analyse pruning for DCGAN models for MNIST which are attacked with TrAIL, ReD, ReX and BAAAN.
and monitor TarDis with respect to attack target and ExpDis with respect to the corresponding compromised DGM for different fractions of pruning.
Figure~\ref{fig:effect-of-pruning} summarises these results.
First we, note that all models except BAAAN are adequately defended by pruning as the Expected Distortion remains low with even 50\% of activations pruned while TarDis increases adequately to distort the attack vector.
We also observe that the 50\%-pruned models for ReD and TraIL only distort the $G(z_\text{trigger})$ and do not recover an MNIST digit.
50\%-pruned ReX on the other hand recovers a digit image.
It is worth noting that ReX introduced additional sparsity within its parameter space to mount the attack and is consequently is a larger model which explains why its expected distortion changes the least with increasing fraction of pruning.
Similarly, BAAAN seems largely unchanged with respect to $G(z_{\text{target}})$ which suggests that the original model might have been overparameterised to begin with.
While this is a preliminary exploration of pruning, we believe that a systematic analysis is warranted for exploring its effectiveness as a defense. 
For instance, investigating the interplay between model capacity, attack algorithms, regularisation schemes (like Dropout) and pruning is a promising next step.

\subsection{Distillation Based Sanitization}
\label{sec:distill-sanitize}
As described in Section~\ref{sec:defenses_recomm} knowledge distillation or fine-tuning approaches offer another line of defense. 
This has been explored previously in~\cite{Aguinaldo2019_Compressing} for compressing DCGANs.
They use a distillation loss that is analogous to the evaluation metric of ExpDis (Section~\ref{sec:attack-analysis}).
It is worth noting that this hasn't been explored for large-scale DGM models. 
Distillation is an expensive process as it requires gradient computation for the entire parameter vector and may not be practical for very large scale models like StyleGAN (with 26M parameters).
Moreover, mean square error or ExpDis might serve as a proxy for small models like DCGAN but its suitability to large scale models is an open question.
And finally, such computation resources might not be readily available at defender's end.

We analyse distillation based sanitization for DCGAN models for MNIST which are attacked with TrAIL, ReD, ReX and BAAAN.
For each model we initialise the student model with the same parameter size as the teacher model and optimise the student model with a distortion based distillation loss.

\begin{figure}
\centering
 \includegraphics[width=0.41\textwidth]{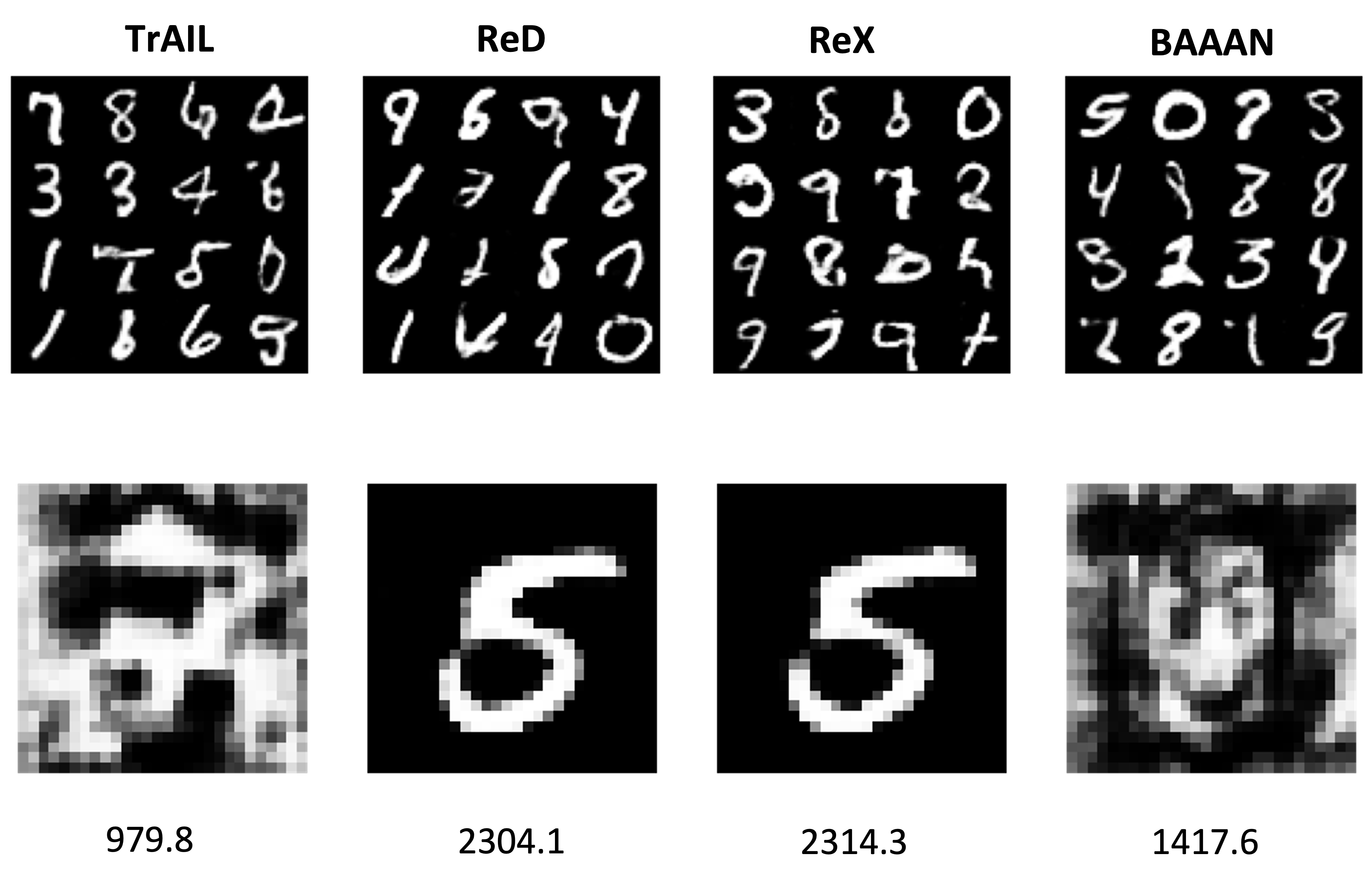}
\caption{Effect of santization on the compromised DCGANs for TrAIL, ReD, ReX and BAAAN - (Upper Row) $G(z)$ for $z\sim P_{\text{sample}}$ and (Lower Row) $G(z_{\text{trigger}})$ with TarDis values.}
\label{fig:effect-of-santisation}
\end{figure}

Figure~\ref{fig:effect-of-santisation} illustrates the results for sanitzed models. 
We observe the following values for (TarDis, ExpDis) respectively for the sanitized models: TrAIL (979.8, 12.219), ReD (2304.1, 17.484), ReX (2314.3, 24.136), and BAAAN (1417.69, 14.395).
Given that TrAIL and BAAAN modify the training algorithm, it is perhaps not surprising that they are the most challenging to santize.
While this approach demonstrates moderate success, as with pruning it requires additional expertise and resources.
Similarly, its interaction with model capacity and suitability of large scale models remain open but encouraging directions of research.




\end{document}